\documentclass[12pt]{iopart}

\usepackage{iopams}
\usepackage{setstack,amssymb,graphicx,color}

\newcommand{\ang}{\theta}
\newcommand{\hmfx}{x}
\renewcommand{\mbox}{\rm}

\begin{document} 

\title{Algebraic damping in the one-dimensional Vlasov equation}

\author{Julien Barr\'e$^1$, Alain Olivetti$^1$ and Yoshiyuki Y Yamaguchi$^2$}
\address{$^1$ Laboratoire J.A. Dieudonn\'e, Universit\'e de Nice Sophia-Antipolis, UMR CNRS 6621, Parc Valrose, F-06108 Nice Cedex 02, France}
\address{$^2$ Department of Applied Mathematics and Physics, Graduate School of Informatics, Kyoto University, Kyoto 606-8501, Japan}
\ead{jbarre@unice.fr}

\begin{abstract}
    We investigate the asymptotic behavior of a perturbation around a
    spatially non homogeneous stable stationary state of a
    one-dimensional Vlasov equation.  Under general hypotheses,
    after transient exponential Landau damping, a perturbation evolving
    according to the linearized Vlasov equation decays algebraically
    with the exponent $-2$ and a well defined frequency.  The
    theoretical results are successfully tested against numerical
    $N$-body simulations, corresponding to the full Vlasov dynamics in
    the large $N$ limit, in the case of the Hamiltonian mean-field
    model.  For this purpose, we use a weighted particles code, which allows us to reduce
    finite size fluctuations and to observe the asymptotic decay in the
    $N$-body simulations.
\end{abstract}
\pacs{05.20.Dd, 45.50.-j, 52.25.Dg, 98.10.+z}
\submitto{\JPA}
\maketitle

\section{Introduction}

Systems of particles interacting through long-range forces are usually
described over a certain time scale by Vlasov equations. This
situation is encountered in various fields of physics: plasma physics,
self-gravitating systems, wave-particles interactions for instance.
One may add here two dimensional fluid dynamics, since the 2D
Euler equation shares many properties with the Vlasov equation.

A Vlasov equation usually admits a continuous infinity of stationary
states. Investigating the stability of these states is a natural
question.  In a celebrated paper \cite{Landau46}, Landau considered
stationary states of a plasma which are homogeneous in space,
and addressed the issue of the asymptotic behavior of a small perturbation through a
Laplace transform analysis of the Vlasov equation linearized around
the stationary state. This was the starting point of an extremely
abundant research on Landau damping in plasma physics, and more
generally on the fate of perturbations around stationary states of the
Vlasov equation, usually homogeneous in space. We focus now on
the known mathematically rigorous results, all of them
obtained in the context of homogeneous stationary solutions of the Vlasov equation. A
rigorous linear treatment ``\`a la Landau'' is provided for instance
in \cite{Maslov85,Degond86}: under very strong regularity hypothesis for the
stationary state and the perturbation (both should be analytic
functions of the velocity), it proves the exponential asymptotic decay
of a solution of the linearized Vlasov equation in a bounded spatial
domain, as predicted by Landau. However, it is known that such a
linear solution may decay at a much slower rate, and even not decay at
all, when the analyticity hypothesis
  for the perturbation (see \cite{Weitzner67,Crownfield77} for references in the physics literature)
or the bounded spatial domain hypothesis \cite{Glassey95} is not satisfied.
Mouhot and Villani proved recently 
 the asymptotic exponential decay of a
perturbation evolving according to the full non linear equation in a
bounded domain, using analytic norms (which implies the analyticity of
the stationary state and its perturbation)~\cite{MouhotVillani,MouhotVillani2}. 
However, it is shown in \cite{Lin10}
(see also \cite{Holloway91} for a previous non rigorous treatment) that such a non linear damping may fail if weaker norms are used.

In many cases, one would like to ask the same question in the case of
non homogeneous stationary solutions. In particular, this is the case
in the astrophysical context, where the concept of Landau damping is
often used \cite{Habib86}. \cite{Maslov85} contains a discussion of the linearized Vlasov equation around a non homogeneous stationary solution. However, the analysis of this situation faces some
technical difficulties. At a formal level, these difficulties may be partially overcome
using the ``matrix'' formulation introduced in the works of Kalnajs \cite{Kalnajs77} and
Polyachenko and Schukhman~\cite{Polyachenko81}. 
Since then, many papers have computed
linear instability rates in an astrophysical context using this method (see \cite{Palmer87,Bertin94}, to
mention just a few). 
Recently, some new methods to investigate the stability and compute such growing 
rates for unstable non homogeneous stationary states have been introduced and tested on toy models~\cite{Jain07,Campa10,Chavanis10,Bachelard10}. Beyond these unstable states, stable oscillating modes in the context of 1D self gravitating models have also been investigated in some details
\cite{Mathur90,Weinberg91}. However, there seems to be very few
analytical studies of what would be the strict analog of Landau
damping: the decay of a perturbation close to a stable non homogeneous
solution of the Vlasov equation. The reason is probably that such a study is technically more 
difficult, since it requires the further step of an analytical continuation. 
Weinberg has performed this step using a numerical approximation and computed the analog of a
Landau damping rate in some cases \cite{Weinberg94} (see also
\cite{Weinberg00}). However, at variance with the homogeneous case, we
shall see that this Landau damping rate never controls the asymptotic
decay of the perturbation.

It is well known that the 2D Euler equation, as well as other related
conservative 2D fluid equations, share a lot of similarities with the
Vlasov equation. There have been many linear investigations of
perturbations around stationary base flows such as a vortex or a shear
flow, starting 
with Rayleigh \cite{Rayleigh79}. Transposing these studies in the 
context of the Vlasov equation, linearized around a non homogeneous stationary state, 
one would expect the following generic picture: the dispersion relation has
branch point singularities on the real axis, which make the analytical
continuation procedure used in Landau's work trickier; if the
stationary state is stable, the asymptotic decay of the perturbation
is controlled by these branch point singularities, and is in general
algebraic. An exponential decay of the perturbation ``\`a la Landau''
may be visible, but it is restricted to an intermediate time window,
before the algebraic decay kicks in.  In particular, it seems reasonable 
to expect that the exponential damping studied by Weinberg
in the gravitational case \cite{Weinberg94} is actually a transient effect 
(this statement does not preclude of course its potential physical relevance).
We remark that such an exponential decay followed by an algebraic decay
has also been studied in a dissipative system of coupled oscillators
\cite{Strogatz};
the decay mechanism in this case is similar to the one in Vlasov systems.

Smereka \cite{Smereka} investigated the asymptotic
behavior of a linearized 1D Vlasov equation around non homogeneous
stationary states, and reached the conclusions outlined above; his 
analysis is however limited to a particular class of interactions, 
and does not contain direct numerical simulations.
Moreover, his conclusion is affected by a non generic choice
of the perturbations, as we will discuss later on.
Besides \cite{Smereka}, we are not aware of 
other studies tackling this problem.

To be more specific, our goals in this paper are
\begin{enumerate}
      \item Study as generally as possible the linear asymptotic
    decay of perturbations around non homogeneous stationary states of
    a 1D Vlasov equation.
      \item Perform explicitly the computations in the simple case of
    the Hamiltonian mean-field (HMF) model.
      \item Compare the analytical results with detailed numerical
    computations; this is made possible by the use of a simple toy
    model such as HMF.
\end{enumerate}

This paper is organized as follows.
All theoretical results are described in Sec.\ref{sec:theory}.
From the linearized Vlasov equation,
we derive two equations for the Fourier-Laplace components
of perturbation and of potential in Sec.\ref{sec:two-equations}.
In Sec.\ref{sec:biorthogonal}
these two equations are formally solved in the Laplace space
with the help of biorthogonal functions; an example of such functions
is given in Sec.\ref{sec:0-2pi} for spatially periodic systems.
The asymptotic dynamics of a perturbation is determined by its singularities 
in the Laplace space; we classify the singularities in Sec.\ref{sec:singularities}.
Focusing on one type of singularities,
which is in many cases the relevant one,
we show that the perturbation asymptotically decays algebraically
with the exponent $-2$ in Sec.\ref{sec:contribution-from-first}.
These general results are applied to the HMF model in Sec.\ref{sec:HMF}; in this case, 
we show that the two components of the magnetization vector
decay with exponents $-3$ and $-2$ respectively.
We note that the exponent $-3$ comes from a special cancellation
due to a symmetry of the HMF model.
The theoretical results for the HMF model
are numerically examined in Sec.\ref{sec:numerics}.
In order to test the two exponents separately, we introduce two types of perturbation
in Sec.\ref{sec:cosine} and in Sec.\ref{sec:sine} respectively.
The last section \ref{sec:conclusion} is devoted to conclusion.

\section{Theory}
\label{sec:theory}

We consider the Vlasov equation in one dimension for the one-particle
distribution function $f(x,p,t)$,
\begin{equation}
    \label{eq:vlasov}
    \partial_{t}f + \partial_{p}H \partial_{x}f - \partial_{x} H \partial_{p}f = 0, 
\end{equation}
where $x \in D\subset\mathbb{R}$ is the position variable,
$p\in\mathbb{R}$ the conjugate momentum variable,
and $H$ is the one-particle Hamiltonian defined by
\begin{equation}
    \label{eq:one-particle-hamiltonian}
    H[f](x,p,t) = \frac{p^{2}}{2} + \phi[f](x,t) + \phi_{\mbox{ext}}(x).
\end{equation}
The potential $\phi[f](x,t)$ is defined from 
the two-body interaction potential $v$ and the distribution $f$ as
\begin{equation}
    \label{eq:potential}
    \phi[f](x,t) = \int_{D} \int_{\mathbb{R}} v(x-y) f(y,p,t) dpdy,
\end{equation}
and $\phi_{\mbox{ext}}(x)$ derives from an external force $F_{\mbox{ext}}(x)$ as
\begin{equation}
    \label{eq:external-force}
    F_{\mbox{ext}}(x)=-\partial_{x}\phi_{\mbox{ext}}(x).
\end{equation}
Although it would be interesting to consider a time dependent 
and/or non-potential external force, we 
focus in this article on a static and potential force such as (\ref{eq:external-force}).
The domain $D$ is typically $\mathbb{R}$ or $[0,2\pi]$:
\begin{enumerate}
      \item {\bf Case~1, $D=\mathbb{R}$:} The model is defined on the
    whole real line. A typical example is a 1D self-gravitating
    system, a very much studied caricature of the more realistic 3D
    self-gravitating systems.
      \item {\bf Case~2, $D=[0,2\pi]$:} The system has periodic
    boundary conditions. Such boundary conditions are sometimes used
    in plasma physics; it is also the setting of the HMF model, a
    paradigmatic toy model for long range interacting systems (see 
    Sec.~\ref{sec:HMF}).
\end{enumerate}
We give in the following subsections a general analysis of the linearized Vlasov equation, 
which applies to the both cases.

\subsection{Two equations to be solved}
\label{sec:two-equations}

Let us call $f_{0}$ a stationary solution of the Vlasov equation
(\ref{eq:vlasov}).  The one-particle Hamiltonian $H[f_{0}](x,p)$ is integrable,
and its trajectories are level curves of
$H[f_0]$. If $D=\mathbb{R}$, one could imagine trajectories that are
unbounded and not periodic. However, the stationarity of $f_0$
imposes that $f_0(x,p)$ is constant along the trajectories; this
imposes that such unbounded non periodic trajectories either do not
exist, or are not populated by the density $f_0$.  As a consequence,
we can always define the angle $\ang$ and the action $J$ from the
original coordinate $(x,p)$.  Strictly speaking, this change of
variables is not always one-to-one: whenever there exists a separatrix
in the one-particle Hamiltonian $H[f_{0}](x,p)$, there are distinct
trajectories with the same value of the action.  Thus, although we
will formally use this change of variables, a careful treatment may be
needed for specific potentials (see for instance~\cite{BOY10}).

We add a perturbation $f_{1}$ to the stationary solution,
and start from an initial condition $f_{0}+f_{1}(t=0)$.
The Hamiltonian $H$ is linear with respect to $f$, and we have
$H[f_{0}+f_{1}]=H[f_{0}]+\phi[f_{1}]$, where
\begin{equation}
    \phi[f_1](x,t)=\phi_1(x,t)=\int_{D} \int_{\mathbb{R}}v(x-y) f_1(y,p,t) dp dy.
    \label{eq:phi1}
\end{equation}
The stationary solution $f_{0}$ must be constant along trajectories of
the Hamiltonian $H[f_0]$; a sufficient condition for this is to take
$f_0$ a function of the action alone. It is not a necessary condition
if there is a separatrix, since two disjoint trajectories correspond
to the same action. Thus, with a slight loss of generality, we will
assume in the following that $f_0$ may be written as $f_0(J)$.  The
unperturbed part of one-particle Hamiltonian is also a function of the
action alone, and written as $H[f_{0}](J)$.  Using the angle-action
variables $(\ang,J)$, we have the linearized Vlasov equation
\begin{equation}
    \partial_t f_1+\Omega(J)\partial_{\ang} f_1- f^{'}_0(J) \partial_{\ang} \phi_1 =0~,
    \label{eq:eqlin}
\end{equation}
where we have defined the frequency $\Omega(J)=dH[f_0]/dJ$. Notice
that the external potential does not enter in this linear equation; it
appears implicitly of course through the definition of the angle-action variables.

To analyze the linearized Vlasov equation (\ref{eq:eqlin}),
we introduce the Fourier-Laplace transform $\hat{u}(m,J,\omega)$ of a 
function $u(\ang,J,t)$ as
\begin{equation}
  \hat{u}(m,J,\omega)
  = \int_{-\pi}^{\pi}d\ang~ e^{-im\ang} \int_{0}^{+\infty}dt~ e^{i\omega t}u(\ang,J,t)
\end{equation}
where $m$ is an integer and ${\mbox Im}(\omega)$ large enough to ensure convergence.
The inverse transform is then
\begin{equation}
  \label{eq:FourierLaplace_inverse}
  u(\ang,J,t)=\frac{1}{(2\pi)^{2}}\sum_{m=-\infty}^{\infty}
  \int_{\Gamma}d\omega~\hat{u}(m,J,\omega) e^{-i\omega t} e^{im\ang}
\end{equation}
where $\Gamma$ is a Bromwich contour running from
$-\infty+i\sigma$ to $+\infty+i\sigma$,
and the real value $\sigma$ is larger
than the imaginary part of any singularity of 
$\hat{u}(m,J,\omega)$ in the complex $\omega$-plane. 

Performing a Fourier transform with respect to $\ang$ and a Laplace
transform with respect to time on (\ref{eq:eqlin}), we obtain,
after simple algebraic manipulations
\begin{equation}
    \label{eq:f1-fourier-laplace}
    \hat{f}_{1}(m,J,\omega) = A(m,J,\omega) \hat{\phi}_{1}(m,J,\omega) + B(m,J,\omega),
\end{equation}
where
\begin{equation}
    \label{eq:A}
    A(m,J,\omega) = \frac{mf^{'}_0(J)}{m\Omega(J)-\omega},
\end{equation}
\begin{equation}
    \label{eq:B}
    B(m,J,\omega) = \frac{g(m,J)}{m\Omega(J)-\omega},
\end{equation}
and $ig(m,J)$ is the Fourier transform of the initial perturbation
$f_1(\ang,J,t=0)$ with respect to $\ang$. We assume
\begin{equation}
    \label{eq:f1t0}
    \int\!\int f_{1}(\ang,J,t=0) d\ang dJ = 0
\end{equation}
and hence $g(0,J)=0$.

The two equations (\ref{eq:phi1}) and (\ref{eq:f1-fourier-laplace})
relate $f_{1}$ to $\phi_{1}$.
The strategy is now to combine these two equations to compute $f_1$ and $\phi_1$.
One sees however that $f_1$ is easily obtained in angle-action variables
whereas $\phi_1$ is more easily expressed in the original $(x,p)$ variables.
To overcome the difficulty of the two natural coordinate basis $(x,p)$ and $(\ang,J)$, 
we follow the standard procedure 
and introduce two families of biorthogonal functions~\cite{CluttonBrock72,Kalnajs77,Polyachenko81}.

\subsection{Biorthogonal functions}
\label{sec:biorthogonal}
We introduce the linear mapping $L_{v}$ by
\begin{equation}
    \label{eq:V}
    L_{v}~:~d\mapsto u
\end{equation}
where $u=L_{v}\cdot d$ is defined by
\begin{equation}
    \label{eq:u}
    u(x) = \int_{D} v(x-y) d(y) dy.
\end{equation}

We assume that there exist two index sets $I$ and $I'$
which satisfy $I'\subset I\subset\mathbb{Z}$,
and two families $\{d_{j}(x)\}_{j\in I}$ and $\{u_{k}(x)\}_{k\in I'}$
which satisfy the following conditions:
\begin{itemize}
      \item[(i)] $\{d_{j}(x)\}_{j\in I}$ is linearly independent 
    and any density function $\rho(x)$ may be expanded as
    \begin{equation}
        \label{eq:d-expansion}
        \rho(x) = \sum_{j\in I} a_{j} d_{j}(x),
    \end{equation}
      \item[(ii)] $\{u_{k}(x)\}_{k\in I'}$ is linearly independent and spans ${\rm Im}(L_{v})$;
    any function $g(x)\in{\rm Im}(L_{v})$ may be expanded as
    \begin{equation}
        \label{eq:u-expansion}
        g(x) = \sum_{k\in I'} b_{j} u_{j}(x),
    \end{equation}
      \item[(iii)] the two families are orthogonal to each other:
    \begin{equation}
        \label{eq:biorthogonal}
        (d_{j},u_{k}) = \int_{D} d_{j}(x) \bar{u}_{k}(x) dx
        = \lambda_{k}\delta_{jk}, ~(j\in I, k\in I')
    \end{equation}
    with $\lambda_{k}\neq 0$,
    and where $\delta_{jk}$ is the Kronecker $\delta$.
      \item[(iv)] For all $k\in I'$, $d_{k}$ and $u_{k}$ satisfy the property $u_k=L_v\cdot d_k$.
    For all $k \in I\setminus I'$, $L_v\cdot d_k =0$.
\end{itemize}
Let us start from the definition of the potential $\phi_{1}(x,t)$ (\ref{eq:phi1}).
This definition may be rewritten as
\begin{equation}
    \label{eq:phi1-rho1}
    \phi_{1}(x,t) = \int_{D} v(x-y) \rho_{1}(y,t) dy
\end{equation}
by using the perturbation density
\begin{equation}
    \label{eq:rho1}
    \rho_{1}(x,t) = \int_{\mathbb{R}} f_{1}(x,p,t) dp.
\end{equation}
From the assumption (i), the perturbation density is expanded in the form
\begin{equation}
    \label{eq:rho1-expansion}
    \rho_{1}(x,t) = \sum_{j\in I} a_{j}(t) d_{j}(x).
\end{equation}
Substituting (\ref{eq:rho1-expansion}) into (\ref{eq:phi1-rho1}) and using (iv),
we obtain an expansion for $\phi_{1}$ in the form
\begin{equation}
    \label{eq:phi1-expansion}
    \phi_{1}(x,t) = \sum_{k\in I'} a_{k}(t) u_{k}(x).
\end{equation}
The Fourier-Laplace transform of (\ref{eq:phi1-expansion}) is expressed by
\begin{equation}
    \label{eq:phi1-fourier-laplace}
    \hat{\phi}_{1}(m,J,\omega) = \sum_{k\in I'} \tilde{a}_{k}(\omega) c_{km}(J),
\end{equation}
where $\tilde{a}_{k}(\omega)$ is the Laplace transform of $a_{k}(t)$,
\begin{equation}
    \label{eq:aj-laplace}
    \tilde{a}_{k}(\omega) = \int_{0}^{\infty} a_{k}(t) e^{i\omega t} dt,
\end{equation}
and $c_{km}(J)$ is the Fourier transform of $u_{k}(x)$,
\begin{equation}
    \label{eq:ckm}
    c_{km}(J) = \int_{-\pi}^{\pi} u_{k}(x) e^{-im\ang} d\ang.
\end{equation}
Substituting (\ref{eq:phi1-fourier-laplace}) into 
(\ref{eq:f1-fourier-laplace}), we obtain
\begin{equation}
    \label{eq:f1-fourier-laplace-expand}
    \hat{f}_{1}(m,J,\omega) = 
    A(m,J,\omega) \sum_{k\in I'} \tilde{a}_{k}(\omega) c_{km}(J)
    + B(m,J,\omega).
\end{equation}
The inverse Fourier transform of (\ref{eq:f1-fourier-laplace-expand})
gives $\tilde{f}_{1}(\ang,J,\omega)$, the Laplace transform of $f_{1}(\ang,J,t)$:
\begin{eqnarray}
    \label{eq:f1-laplace}
         \tilde{f}_{1}(\ang,J,\omega)
        &=& \frac{1}{2\pi} \sum_{m\in\mathbb{Z}} \hat{f}_{1}(m,J,\omega) e^{im\ang} \\
         &=& \frac{1}{2\pi} \sum_{m\in\mathbb{Z}}
        \left[ A(m,J,\omega) \sum_{k\in I'} \tilde{a}_{k}(\omega) c_{km}(J)
          +B(m,J,\omega) \right] e^{im\ang}. 
\end{eqnarray}
We observe that $\tilde{f}_{1}(\ang,J,\omega)$, and hence $f_{1}(\ang,J,t)$,
are determined by the $\{\tilde{a}_{k}(\omega)\}_{k\in I'}$,
so that the subset $\{\tilde{a}_{j}(\omega)\}_{j\in I\setminus I'}$ is not necessary.
We therefore seek a solution for the subfamily $\{\tilde{a}_{k}(\omega)\}_{k\in I'}$
instead of the whole family $\{\tilde{a}_{j}(\omega)\}_{j\in I}$.
For this purpose, we multiply (\ref{eq:f1-laplace}) by $\bar{u}_{l}(x)~(l\in I')$ 
and integrate it over $\ang$ and $J$.
Using the biorthogonality relation~(\ref{eq:biorthogonal})
and noting that the change of variable $(x,p)\mapsto (\ang,J)$ is symplectic
$dx\wedge dp=d\ang\wedge dJ$, the left-hand-side of (\ref{eq:f1-laplace}) becomes
\begin{eqnarray}
    \label{eq:f1-laplace-lhs}
         \int\int \tilde{f}_{1}(\ang,J,\omega) \bar{u}_{l}(x) d\ang dJ 
        &=& \int_{\mathbb{R}}\int_{D} \tilde{f}_{1}(x,p,\omega) \bar{u}_{l}(x) dxdp\\
        &=& \int_{D} \tilde{\rho}_{1}(x,\omega) \bar{u}_{l}(x) dx
        = \tilde{a}_{l}(\omega) \lambda_{l}.
\end{eqnarray}
To derive the last equality, we have used the fact
that the Laplace transform of $\rho_{1}(x,t)$ (\ref{eq:rho1-expansion}) is
\begin{equation}
    \label{eq:rho1-laplace}
    \tilde{\rho}_{1}(x,\omega) = \sum_{j\in I} \tilde{a}_{j}(\omega) d_{j}(x).
\end{equation}
On the other hand, the right-hand-side of (\ref{eq:f1-laplace}), 
submitted to the same operations, becomes
\begin{eqnarray}
    \label{eq:f1-laplace-rhs}
         \int\!\int \frac{1}{2\pi} \sum_{m\in\mathbb{Z}} \Big[
          A(m,J,\omega)
          \sum_{k\in I'}\tilde{a}_{k}(\omega) c_{km}(J)+ B(m,J,\omega) \Big] \bar{u}_{l}(x)  e^{im\ang} d\ang dJ \nonumber \\
         = \frac{1}{2\pi} \sum_{m\in\mathbb{Z}} \sum_{k\in I'}
        \tilde{a}_{k}(\omega) \int 
        A(m,J,\omega) c_{km}(J) \bar{c}_{lm}(J) dJ \nonumber \\ 
         + \frac{1}{2\pi} \sum_{m\in\mathbb{Z}} \int
        B(m,J,\omega) \bar{c}_{lm}(J) dJ.
\end{eqnarray}
Remembering the definitions of $A(m,J,\omega)$ and $B(m,J,\omega)$,
we introduce the functions
\begin{equation}
    \label{eq:F}
    F_{lk}(\omega) = \frac{1}{2\pi} \sum_{m\in\mathbb{Z}} \int
    \frac{mf^{'}_0(J)}{m\Omega(J)-\omega}
    c_{km}(J) \bar{c}_{lm}(J) dJ,  \quad l,k\in I'
\end{equation}
\begin{equation}
    \label{eq:G}
    G_{l}(\omega) = \frac{1}{2\pi} \sum_{m\in\mathbb{Z}} \int
    \frac{g(m,J)}{m\Omega(J)-\omega} \bar{c}_{lm}(J) dJ, \quad l\in I'
\end{equation}
where contributions from $m=0$ vanish not only for $F_{lk}$
but also for $G_{l}$, thanks to assumption (\ref{eq:f1t0}).
We further define the $(\sharp I')\times (\sharp I')$ matrices
$\Lambda$ and $F(\omega)=(F_{lk}(\omega))_{l,k\in I'}$,
where $\Lambda$ is diagonal with elements $\{\lambda_{k}\}_{k\in I'}$,
and the $(\sharp I')$-dimensional vectors
$G(\omega)=(G_{l}(\omega))_{l\in I'}$ and 
$\tilde{a}(\omega)=(\tilde{a}_{k}(\omega))_{k\in I'}$.
Using the above matrices and vectors,
the equation for $\tilde{a}(\omega)$ reads in matrix form:
\begin{equation}
    \label{eq:a-determine}
    [ \Lambda - F(\omega) ] \tilde{a}(\omega) = G(\omega).
\end{equation}
The equation (\ref{eq:a-determine}) is formally solved as
\begin{equation}
    \label{eq:a-formalsol}
    \tilde{a}(\omega) = [\Lambda - F(\omega)]^{-1} G(\omega),
\end{equation}
and the temporal evolution of the $\{a_{k}(t)\}_{k\in I'}$, and of $f_{1}$,
is obtained from the inverse Laplace transform of $\tilde{a}(\omega)$.
The inverse matrix $[\Lambda-F(\omega)]^{-1}$ does not always exist
since the determinant $\det(\Lambda - F(\omega))$ is not always non-zero.
This determinant is sometimes called the dispersion function,
and its roots are poles of $\tilde{a}(\omega)$.
We will discuss the singularities of $\tilde{a}(\omega)$ 
in Sec.\ref{sec:singularities}
after giving an example of the two families $\{d_{j}\}_{j\in I}$ and
$\{u_{k}\}_{k\in I'}$ in the domain $D=[0,2\pi]$.

\subsection{Example: $D=[0,2\pi]$}
\label{sec:0-2pi}
If we consider the domain $D=[0,2\pi]$ with periodic boundary condition,
the interaction potential $v(x)$ must be also $2\pi$-periodic and is expanded
in Fourier series as
\begin{equation}
    \label{eq:v-fourier}
    v(x) = \frac{1}{2\pi} \sum_{m\in\mathbb{Z}} v_{m} e^{imx},
\end{equation}
where the coefficients $v_{m}$ are determined by
\begin{equation}
    \label{eq:vm}
    v_{m} = \int_{-\pi}^{\pi} v(x) e^{-imx} dx.
\end{equation}
We can choose the two families $\{d_{j}(x)\}_{j\in I}$
and $\{u_{k}(x)\}_{k\in I'}$ as
\begin{equation}
    \label{eq:dj-periodic}
    d_{j}(x) = \frac{1}{2\pi} e^{ijx}, \quad j\in I=\mathbb{Z},
\end{equation}
and
\begin{equation}
    \label{eq:uk-periodic}
    u_{k}(x) = \int_{0}^{2\pi} v(x-y) d_{k}(y) dy = \frac{1}{2\pi} v_{k} e^{ikx},
    \quad k\in I'
\end{equation}
where the index set $I'$ is defined by
\begin{equation}
    \label{eq:Idash}
    I' = \{ k\in\mathbb{Z} ~|~ v_{k}\neq 0\}.
\end{equation}
From the definition of $u_{k}(x)$ and $I'$
the assumption (iv) is satisfied.
Expansions on the two families $\{d_{j}(x)\}_{j\in\mathbb{Z}}$
and $\{u_{k}(x)\}_{k\in I'}$ are essentially Fourier expansions,
and hence the assumptions (i)-(iii) are also satisfied.
The factors $\lambda_{k}$ are $\lambda_{k}=v_{k}/2\pi$ for $k\in I'$.
Generically, all $v_k$ are non zero, and $I=I'=\mathbb{Z}$.
However, for the HMF model, which we will introduce in section~\ref{sec:HMF}, 
$v(x)=-\cos x$ and ${\mbox Im}(L_{v})$ is $2$-dimensional.
Accordingly the index set $I'$ is $I'=\{1,-1\}$
and the matrix $\Lambda-F(\omega)$ in (\ref{eq:a-determine})
is a $2\times 2$ matrix.

\subsection{Singularities of $\tilde{a}(\omega)$}
\label{sec:singularities}

From  the  knowledge of  the  functions  $\tilde{a}(\omega)$, one  may
easily  compute  the  time  evolution  of the  potential  and  density
perturbations, through an inverse Laplace transform. The asymptotic in
time behavior of this inverse Laplace transform will be determined by
the singularities of the functions  $\tilde{a}_{k}(\omega)~(k\in I')$
in the complex plane. We turn now to the study of these singularities.

The matrix coefficients $F_{lk}(\omega)$ and the vector coefficients $G_{l}(\omega)$
are defined through integrals over the real variable $J$,
see (\ref{eq:F}) and (\ref{eq:G}).
These integrals are naturally defined in the whole half
plane ${\mbox Im}(\omega)>0$. Note however, that expressions (\ref{eq:F})
and (\ref{eq:G}) are in general not
properly defined for $\omega \in \mathbb{R}$, as in this case
$m\Omega(J)-\omega$ may vanish. For ${\mbox Im}(\omega)\leq 0$, we will
actually have to consider rather the analytical continuations of the
expressions~(\ref{eq:F}) and (\ref{eq:G}).
We may have two kinds of singularities, described in the following.

The first kind of singularities is poles,
coming from roots of the dispersion function $\det(\Lambda-F(\omega))$.
If such a root exists in the half plane ${\mbox Im}(\omega)>0$,
it corresponds to an eigenvalue of the linearized Vlasov operator,
and it yields an exponential growth of the perturbation.
In the upper half plane ${\mbox Im}(\omega)>0$, the only possible
singularities for $\tilde{a}(\omega)$ are the poles.
We assume in the following that the reference stationary state $f_0$ is linearly stable,
so that the determinant of $\Lambda-F(\omega)$ does not have any roots with
${\mbox Im}(\omega)>0$.
If $\det(\Lambda-F(\omega))$ has a root on the real axis ${\mbox Im}(\omega)=0$,
it corresponds to a purely oscillating mode.
This is compatible with a linearly stable $f_0$, but corresponds to a
non decaying perturbation; we also assume in the following that this
does not happen.  The analytical continuation of $\det(\Lambda-F(\omega))$
in the lower half plane ${\mbox Im}(\omega)<0$ may also have roots.
They correspond to ``Landau poles'' for $\tilde{a}(\omega)$,
and give rise to an exponential damping of the perturbation.
This damping behavior is known as Landau damping,
and has been studied in the gravitational case in \cite{Weinberg94,Weinberg00},
and in the HMF case in \cite{BOY10}.
As anticipated in the introduction, we will see that this exponential
damping, if it exists, is subdominant in the large time region.

The second kind of singularities comes from the integral in
(\ref{eq:F}) and (\ref{eq:G}),
and appears on the real axis of ${\mbox Im}(\omega)=0$.
We have to study the singularities of functions
\begin{equation}
    \varphi(z)=\int_a^b \frac{\psi(J)}{m\Omega(J)-z}dJ
    \label{eq:varphi}
\end{equation}
properly defined for ${\mbox Im}(z)>0$,
where $\psi$ and $\Omega$ are real functions,
and $[a,b]\subset\mathbb{R}\cup\{+\infty\}$.
$J$ has to be thought of as the action coordinate,
and $\Omega$ as the associated frequency.
We assume that $\psi$ is analytic.
We may set $m\neq 0$ since the contributions from $m=0$
in (\ref{eq:F}) and (\ref{eq:G}) vanish.
We now show that $\varphi(z)$ is regular for $z\in\mathbb{R}$
except for special points, and will classify the special points into three types.
Notice that both functions $F_{kl}$ and $G_{k}$ fit in this framework.

If the equation $m\Omega(J)=z$ has no solutions in $[a,b]$ for any $z$
in a neighborhood of real $z_{0}$,
then $\varphi$ is analytic in a neighborhood of $z_0$.
Assume now that the equation $m\Omega(J)=z$ has one or several branches
of solutions $J_{i}^{\ast}(z)\in [a,b]$ in a neighborhood of $z_{0}$,
where all $J_{i}^{\ast}(z)\in [a,b]$ are regular as functions of $z$.
In this case, $\varphi$ can be analytically continued from the open half plane
${\mbox Im}(z)>0$ to the neighborhood of $z_{0}$
by taking into account the possible residue contributions of the roots $J_{i}^{\ast}(z)$,
in a straightforward generalization of the ``Landau prescription''.
Thus, generically, no singularity of $\varphi$ appears at $z=z_0$.

\begin{figure}[ht]
    \centering
    \includegraphics[width=8cm]{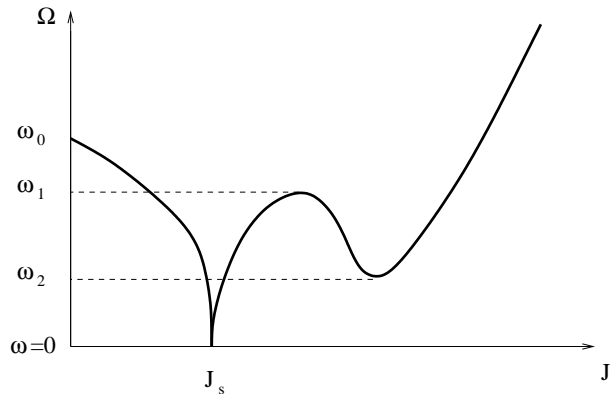}
    \caption{An example of function $\Omega(J)$, with the special points
      at the origin of singularities for the associated function $\varphi$
      defined as in Eq.~(\ref{eq:varphi}). \label{fig:Omega}}
\end{figure}

The singularities of $\varphi$ are hence associated with special points $z_0$,
such that the branches of solutions $J_i^\ast(z)\in [a,b]$ of
the equation $m\Omega(J)=z$ undergo a bifurcation or are
singular. This may happen in the following three types,
illustrated on Fig.~\ref{fig:Omega}:
\begin{enumerate}
      \item $z_0=m\Omega(a)$
    or $z_0=m\Omega(b)$, when one or both are
    finite. This is a common situation, generically encountered in 1D
    self-gravitating systems \cite{Mathur90} as well as in the HMF
    model \cite{BOY10}.
    For instance, this mechanism creates a singularity around
    the frequency $m\omega_{0}=m\Omega(J=0)$, where $J=0$
    corresponds to the minimum of the effective potential $\phi[f_0]+\phi_{\mbox{ext}}$,
    see Fig.~\ref{fig:Omega}.
    We will show in Sec.\ref{sec:contribution-from-first}
      that this singularity is logarithmic.
      \item $z_0$ corresponds to a $J$ such that $\Omega(J)$ is
    singular: this may be an action $J_{s}$ corresponding to a separatrix;
    an illustration is given on Fig.~\ref{fig:Omega}, for $J=J_s$,
    $\omega=0$.  This is not a common situation for the 1D
    self-gravitating models we have in mind. This generically happens
    however for periodic systems, where the trajectories in the
    one-particle Hamiltonian $H[f_{0}]$ may be oscillating or librating,
    and the two regions are delimited by a separatrix.  We will
    see such a situation in the HMF case.
      \item $z_0$ corresponds to a $J$ such that $\Omega^{'}(J)=0$:
    this corresponds generically to a local maximum or minimum of the
    frequency (see frequencies $\omega_1$ and $\omega_2$ on
    Fig.~\ref{fig:Omega}).  We are not aware of any model studied in
    the literature where this phenomenon happens. This is
    certainly an interesting case to study, especially in view of the
    results obtained in an analogous situation for the diocotron instability of a magnetically confined electron column \cite{SmithRosenbluth90}, and recently for the 2D Euler
    equation \cite{BouchetMorita09}. We will not be concerned with
    this type of singularities in the following.
\end{enumerate}

  In the next subsection we estimate the asymptotic relaxation of $a_{k}(t)$
  by considering contributions from singularities of the first type,
  since they should be in many cases the relevant singularities.
  We will confirm whether the estimation is valid
  by performing direct $N$-body simulations in Sec.\ref{sec:numerics}.

\subsection{Contribution from singularities of the first type}
\label{sec:contribution-from-first}

We concentrate now on the singularities of the first type at
$z=m\Omega(a)=m\omega_{a}$.
Notice that if $b=+\infty$,
and $\omega_{\infty}=\Omega(+\infty)$ is finite,
$z=m\omega_{\infty}$ is a singular point;
indeed, the number of solutions of the equation $m\Omega(J)=z$
changes from $z>m\omega_{\infty}$ to $z<m\omega_{\infty}$.
We expect this singularity to be irrelevant if $f_{0}(J)$ decreases rapidly enough
as $J\to\infty$, and we neglect it in the following.

To analyze the function $\varphi(z)$ (\ref{eq:varphi})
around $z=m\omega_{a}$,
we expand $\Omega(J)$ and $\psi(J)$ in power series around $J=a$ as
\begin{equation}
    \label{eq:Omega-expand}
    \Omega(J) = \omega_a + c(J-a)+O\left( (J-a)^2 \right)
\end{equation}
and
\begin{equation}
    \label{eq:psi-expand}
    \psi(J)=d(J-a)^{\nu}+o\left( (J-a)^{\nu} \right)
\end{equation}
for some $\nu\in \mathbb{N}$. 
The exponent $\nu$ will be determined for a given $m$ later.
Substituting these assumptions into $\varphi$ 
and changing $J-a$ to $J$,
this yields the following singular part for $\varphi$:
\begin{equation}
    \label{eq:varphi-expand}
    \varphi(z) = C \int_{0}^{b-a} \frac{J^{\nu}}{J-\zeta} dJ +~{\mbox (regular~part)} 
\end{equation}
with the constant $C=d/mc$ and $\zeta=(z-m\omega_{a})/mc$.
Using the equality
\begin{equation}
    \label{eq:varphi-equality}
    \frac{J^{\nu}}{J-\zeta} = J^{\nu-1} + \zeta \frac{J^{\nu-1}}{J-\zeta} 
\end{equation}
recursively, the singular part of $\varphi(z)$,
which comes from the lower bound of the integral (\ref{eq:varphi-expand}),
behaves as a logarithm times a power around $\zeta=0$:
\begin{equation}
    \varphi(z) = {\mbox cste} \times (z-m\omega_a)^{\nu} \ln(z-m\omega_a)
    +{\mbox (regular~part)}.
    \label{eq:singu}
\end{equation}
Thus, at $\omega=m\omega_{a}$, the matrix elements $F_{kl}(\omega)$
and the vector elements $G_{k}(\omega)$ have singularities of the type
$(\omega-m\omega_a)^{\nu} \ln(\omega-m\omega_a)$.
From (\ref{eq:a-formalsol}), $\tilde{a}_k(\omega)$ is expressed as a sum, 
product and ratio of $F_{kl}(\omega)$ and $G_{k}(\omega)$ functions. Thus, the
leading singularity of $\tilde{a}_k(\omega)$ at $\omega=m\omega_{0}$ is also
of the type $(\omega-m\omega_a)^{\nu} \ln(\omega-m\omega_a)$ for some $\nu$.
$a_k(t)$ is the inverse Laplace transform of $\tilde{a}_k(\omega)$. Since $\tilde{a}_k$ has only logarithmic singularities on the real axis, we may deform the contour $\Gamma$ of the inverse Laplace transform down to the real axis.
The inverse Laplace transform then becomes an inverse Fourier transform.

The asymptotic decay of $a_{k}(t)$ is then determined by the strongest singularity
of $\tilde{a}_k(\omega)$ on the real axis (see \cite{Lighthill} p.52).
A singularity such as (\ref{eq:singu}) yields an asymptotic decay as (see \cite{Lighthill} p.42)
\begin{equation}
    \label{eq:akt-estimate}
    a_k(t)\sim {\mbox cste} \frac{e^{-im\omega_a t}}{t^{\nu+1}}.
\end{equation}
See \ref{sec:estimate} for a heuristic explanation on how to obtain estimates such 
as~(\ref{eq:akt-estimate}).

To obtain $\nu$ for a given $m$,
we need to introduce some assumptions about the system we consider.
We assume that the potential created by the stationary state
(interaction potential + external potential $\phi[f_{0}](x)+\phi_{\mbox{ext}}(x)$) has a single minimum at $J=a=0$,
and is quadratic with respect to $x$
around its minimum at leading order, except for an irrelevant constant term.
This is the case in many situations of
interests, such as stationary states for a 1D self-gravitating system.
We also assume that $f_0(J)$ is analytic, and decays fast enough at
infinity (for instance exponentially). This assumption excludes for
instance truncated $f_0$, or compactly supported stationary states. As an
example, the thermal equilibria of a 1D self-gravitating system
or of the HMF model satisfy all assumptions.
Finally, we assume that the perturbation is also analytic.

Under the above assumptions, we now estimate the exponent $\nu$.
From (\ref{eq:F}) and~(\ref{eq:G}), we see that the function $\psi(J)$ reads
\begin{displaymath}
    \psi(J) = f'_{0}(J) c_{km}(J) \bar{c}_{lm}(J)
\end{displaymath}
for $F_{kl}(\omega)$ functions and
\begin{displaymath}
    \psi(J) = g(m,J) \bar{c}_{lm}(J)
\end{displaymath}
for $G(\omega)$ functions; we now expand these functions with respect to $J$.
Let us start from the function $c_{km}(J)$.
In the limit of $J\to 0$, 
motion is harmonic with the frequency $\omega_{0}$,
and the position $x$ is written in polar coordinates
using angle-action variables as $x\propto J^{1/2}\sin\ang$.
Expanding $u_{k}(x)$ with respect to $x$,
and substituting the above expression of $x$ into the expansion,
the function $c_{km}(J)$ reads
\begin{displaymath}
    c_{km}(J)\sim \int_{0}^{2\pi} \sum_{n=0}^{\infty} \frac{u_{k}^{(n)}(0)}{n!} J^{n/2}\sin^{n}\ang~e^{-im\ang}d\ang,
\end{displaymath}
where $u_{k}^{(n)}$ is the $n$-th derivative of $u_{k}$.
The first non-vanishing term corresponds to $n=|m|$,
and hence the leading order for $c_{km}(J)$ in a small $J$ expansion is
\begin{equation}
    \label{eq:ckm-leading}
    c_{km}(J) \sim J^{|m|/2}.
\end{equation}
$c_{km}(J)$ is the Fourier transform of $u_{k}(x)$; in a similar way, 
the leading order in a small $J$ expansion for $g(m,J)$, the Fourier transform of 
the initial perturbation, is
\begin{equation}
    \label{eq:gm-leading}
    g(m,J) \sim J^{|m|/2},
\end{equation}
since the perturbation is regular.
The function $f'_{0}(J)$ is regular and hence the leading order is constant.
Consequently, the function $\psi(J)$ is, at leading order
\begin{equation}
    \label{eq:psi-leading}
    \psi(J) \sim J^{|m|}
\end{equation}
both for $F$ and $G$ functions. Hence, for each term in the infinite series
defining the coefficient $F_{kl}(\omega)$ (\ref{eq:F}) and $G_l(\omega)$ (\ref{eq:G}), 
we have $\nu=|m|$.

Going back to (\ref{eq:F}), (\ref{eq:G}) and making use of
section~\ref{sec:singularities}, we see that the strongest
singularities for $F_{kl}(\omega)$ and $G_l(\omega)$ come from the
$m=\pm 1$ terms in the sum over $m$, so that the exponent is $\nu=1$.
We conclude using (\ref{eq:akt-estimate})
 that the functions $a_{k}(t)$, 
under the hypothesis of this section, decay as $e^{-i\omega_0 t}/t^{2}$.

This result has to be compared with \cite{Smereka}, which finds a
decay exponent $3/2$. Since we have performed the same kind of
analysis as Smereka does in \cite{Smereka}, this discrepancy is surprising
even if the class of Hamiltonians studied is different.
It may be traced back to the fact that this author uses the following hypothesis for
the perturbation: 
$\lim_{J\to 0}g(m,J)\neq 0 $ even for $m\neq 0$;
this would correspond to a singular perturbation,
since the initial perturbation $f_{1}$ is not well-defined
in the limit $J\to 0$, because
$\lim_{J\to 0}e^{im\ang}g(m,J)\neq \lim_{J\to 0}e^{im\ang'}g(m,J)$
as soon as $e^{im(\ang-\ang')}\neq 1$.
This singular perturbation implies
that $g(m,J)\sim J^{0}$  in the limit  $J\to 0$ instead of (\ref{eq:gm-leading}),
and hence $\psi(J)=g(m,J)\bar{c}_{lm}(J)\sim J^{|m|/2}$.
Accordingly, the strongest singularities for $G_{l}(\omega)$ (\ref{eq:G}), 
coming from the $m=\pm 1$ terms, corresponds to $\nu=1/2$.
Using considerations similar to the ones described in \ref{sec:estimate}
Smereka showed that this gives a decay exponent $3/2$.
Considering in Smereka's setting a regular perturbation, as is more
natural (and as actually assumed in Eq. (21) of \cite{Smereka}), would
yield the $-2$ exponent also for the class of Hamiltonians studied in
\cite{Smereka}.

We may also compare this result to the asymptotic decay of
perturbations around a stationary 2D shear flow or a vortex. In these
cases, the longitudinal (resp. transverse) velocity perturbation
asymptotically decays as $1/t$ (resp. as $1/t^2$), without temporal
oscillations.

\subsection{Example of the HMF model}
\label{sec:HMF}
In this section, we analyze in more details a specific example,
the HMF model whose Hamiltonian is 
\begin{equation}
    \label{eq:hmf}
    H(\hmfx,p) = \sum_{j=1}^{N} \frac{p_{j}^{2}}{2}+\frac{1}{2N}
    \sum_{j=1}^N\sum_{k=1}^N[1-\cos(\hmfx_j-\hmfx_k)].
\end{equation}
The canonical equation of motion of the HMF model
is described through the magnetization ${\bf M}=(M_{x},M_{y})$ defined by
\begin{equation}
    \label{eq:MxMy0}
    M_{x} = \frac{1}{N} \sum_{j=1}^{N} \cos\hmfx_{j}, \quad
    M_{y} = \frac{1}{N} \sum_{j=1}^{N} \sin\hmfx_{j},
\end{equation}
and hence the computational cost is $O(N)$ for each time step,
although the number of interactions between the $N$ particles is $O(N^{2})$.
This advantage allows precise numerical tests of the predictions.
The associated Vlasov equation reads:
\begin{equation}
  \frac{\partial f}{\partial t}
  + p \frac{\partial f}{\partial\hmfx}
  - \frac{\partial \phi[f]}{\partial\hmfx}
  \frac{\partial f}{\partial p} = 0~,
\end{equation}
with
\begin{equation}
  \label{eq:potential-M}
  \phi[f](\hmfx,t) = - M_x[f] \cos\hmfx- M_y[f] \sin\hmfx
\end{equation}
and
\begin{eqnarray}
    \label{eq:MxMy}
         M_{x}[f](t) &=& \int_{-\infty}^{\infty} \int_{0}^{2\pi} \cos\hmfx f(\hmfx,p,t) d\hmfx dp, \\
         M_{y}[f](t) &=& \int_{-\infty}^{\infty} \int_{0}^{2\pi} \sin\hmfx f(\hmfx,p,t) d\hmfx dp. 
\end{eqnarray}
Note that this is a pendulum potential, so that the dynamics admits a separatrix.
The action-angle variables are explicitly written in terms of elliptic
integrals~\cite{BOY10}.
Without loss of generality, we consider a stationary solution with $M_y=0$, and
write $M^{(1)}_x(t)$ and $M^{(1)}_y(t)$ the magnetization perturbations.

For convenience,
we choose real functions for the family $\{u_{k}\}_{k\in I'}$:
$u_{c}(\hmfx)=\cos\hmfx$ and $u_{s}(\hmfx)=\sin\hmfx$ and $I'=\{c,s\}$.
The coefficients $a_{c}(t)$ and $a_{s}(t)$ of the potential $\phi_{1}(x,t)$,
\begin{displaymath}
    \phi_{1}(x,t) = -a_{c}(t)\cos\hmfx - a_{s}(t)\sin\hmfx,
\end{displaymath}
correspond to $M^{(1)}_{x}(t)$ and $M^{(1)}_{y}(t)$ respectively.

The expansions of $u_c$ and $u_s$ in the Fourier series of the angle variable $\theta$ define
the coefficients $\{c_{cm}(J)\}_{m\in\mathbb{Z}}$ and $\{c_{sm}(J)\}_{m\in\mathbb{Z}}$
as in~(\ref{eq:ckm}); to simplify the notations, we rename these coefficients
$c_m(J)$ and $s_m(J)$ respectively. 

This choice for the family $\{u_k\}$ makes the matrix $\Lambda-F(\omega)$ diagonal~\cite{BOY10}, with
\begin{eqnarray}
    F_{cc}(\omega) &=& \frac{1}{2\pi} \sum_{m\in\mathbb{Z}} \int
    \frac{mf^{'}_0(J)}{m\Omega(J)-\omega}
    |c_{m}(J)|^2  dJ ,
\label{eq:Fcc}\\
F_{ss}(\omega) &=& \frac{1}{2\pi} \sum_{m\in\mathbb{Z}} \int
    \frac{mf^{'}_0(J)}{m\Omega(J)-\omega}
    |s_{m}(J)|^2  dJ ,\\
F_{cs}(\omega)&=&F_{sc}(\omega)=0 .
\end{eqnarray}
We also have the following expressions for $G$:
\begin{eqnarray}
    G_{c}(\omega) &=& \frac{1}{2\pi} \sum_{m\in\mathbb{Z}} \int
    \frac{g(m,J)}{m\Omega(J)-\omega} \bar{c}_{m}(J) dJ ,
\label{eq:Gc}\\
 G_{s}(\omega) &=& \frac{1}{2\pi} \sum_{m\in\mathbb{Z}} \int
    \frac{g(m,J)}{m\Omega(J)-\omega} \bar{s}_{m}(J) dJ. 
\end{eqnarray}

{\it A priori}, according to the general discussion in the previous section, the leading 
singularity of $F_{cc}$ and $F_{ss}$ is located at $\omega=\omega_{0}=\sqrt{M_{0}}$,
and has an index $\nu=1$, since $\omega_{0}$ here
plays the role of $\omega_{a}$ in the previous section \ref{sec:contribution-from-first}.
However, due to the symmetries of the system, a further cancellation occurs:
the function $c_m(J)$ identically vanishes for all $m$ odd and all $J<J_s$, with $J_s$ 
the action at the separatrix. Thus, the strongest singularity for $F_{cc}$ and $G_c$ 
actually comes from the $m=2$ term in (\ref{eq:Fcc}) and (\ref{eq:Gc}), and is located 
at $\omega=2\omega_0$; its index is $\nu=2$.  This has an interesting consequence on 
the asymptotic behavior of $M^{(1)}_x(t)$: since it is now governed by a singularity with 
index $\nu=2$, we expect
\begin{equation}
    \label{Eq:HMF_Mx_decay_prediction}
    M^{(1)}_x(t)\sim \frac{e^{-2i\omega_0t}}{t^3}~{\mbox for}~t\to\infty
\end{equation}
at variance with $M^{(1)}_y(t)$, which is still governed by a $\nu=1$ singularity
\begin{equation}
    \label{Eq:HMF_My_decay_prediction}
    M^{(1)}_y(t)\sim \frac{e^{-i\omega_0t}}{t^2}~{\mbox for}~t\to\infty.
\end{equation}
This feature makes the HMF model particularly suitable for a numerical test of 
the theory developed in this section, as we should be able to probe two different 
asymptotic behaviors for $M^{(1)}_x(t)$ and $M^{(1)}_y(t)$.

\section{Numerical simulations}
\label{sec:numerics}

In this section we test numerically the linear predictions of the
previous section, on the example of the HMF model, by solving the
whole (non linear) Vlasov equation. Solving the Vlasov equation over
long times may be a very heavy numerical task, or even impossible with
current computers. In this case, several features help: the model is
one dimensional, it is particularly simple, and we only need to solve
the Vlasov equation close to a stationary state.

A natural strategy could be to solve directly the Hamiltonian $N$-body
dynamics, with $N$ large enough; we know that this provides an
approximation to the continuous Vlasov evolution. We have found that
the finite-$N$ fluctuations were too big to allow a test of the
asymptotic in time regime.  Another strategy would be to use a
standard Vlasov solver; for instance, a semi-Lagrangian method has
already been used for HMF\cite{buyl10}. This resulted in very heavy
computations.  Finally, we have chosen to introduce an algorithm
relying on a Hamiltonian simulation of appropriately weighted particles
\cite{WollmanOzizmir96},
it provides a very convenient tool to test the theoretical predictions.
The algorithm is described and discussed in \ref{sec:semi-vlasov-code}.
We have tested the weighted particles 
algorithm against (i)~a semi-lagrangian code (ii)~ a simple unweighted $N$-body code (results not reported). The temporal evolution of the magnetization from the three codes are in good agreement up to a certain time. The unweighted $N$-body simulation becomes dominated by finite size fluctuations much earlier than the weighted particles' one.

All simulations discussed in the following were performed using the
weighted particles algorithm, close to a thermal equilibrium stationary
state, parametrized by the temperature $T=1/\beta$:
\begin{equation}
f_0(\hmfx,p)=\mathcal{N}e^{-\beta\left(p^2/2-M_0\cos\hmfx\right)}
\end{equation}
where $\mathcal{N}$ is the normalization and the magnetization
$M_0(T)$ is solution of a consistency equation~\cite{Inagaki93b}.  The
thermodynamical equilibrium state of the HMF is non homogeneous (that is
$M_0\neq 0$) as soon as $T<0.5$.  In the following we only use
$T=0.1$; larger temperatures resulted in increased fluctuations and
made it more difficult to reach the asymptotic in time regime.
The magnetization is $M_{0}=0.946$
and the harmonic frequency is $\omega_{0}=\sqrt{M_{0}}=0.972$ for $T=0.1$.

\subsection{Cosine perturbation}
\label{sec:cosine}

We consider first a cosine perturbation of the thermal equilibrium:
\begin{equation}
    \label{Eq:Thermal_equilibrium_plus_Cosine_perturbation}
    f_0(\hmfx,p)+f_{1}(\hmfx,p)
    =\mathcal{N}_ae^{-\beta\left(p^2/2-M_0\cos\hmfx\right)}\left(1+a\cos\hmfx\right),
\end{equation}
where $|a|$ is small enough.
This perturbation is compatible with the $(\hmfx,p) \to (-\hmfx,-p)$ symmetry of the canonical equation of motion,
and $M_{y}^{(1)}(t)$ is then identically equal to zero.
We may restrict the initial points such that $p_{i}(0)>0$,
and obtain the temporal evolutions of particles
which are initially in the lower half of $\mu$ space
without direct computations, using this symmetry.

\begin{figure}[!ht]
    \begin{center}
        \includegraphics[scale=0.275]{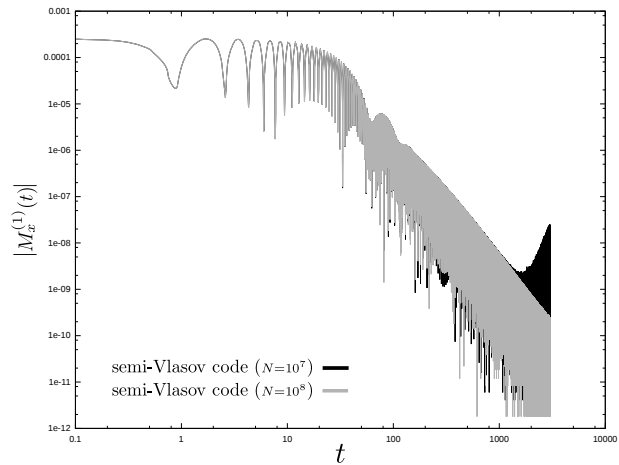}
        \caption{Temporal evolution of $M_x^{(1)}(t)$
          for an initial condition as
          in~(\ref{Eq:Thermal_equilibrium_plus_Cosine_perturbation}).
          We used the weighted particles code with $T=0.1$, $a=0.1$,
          $N=10^{7}$ (black curve) and $N=10^{8}$ (gray curve).  The
          initial weighted points are equally distributed in
          $]-\pi,\pi]\times[-3,3]$, and we used the
          $(\hmfx,p)\to (-\hmfx,-p)$ symmetry in order to reduce the computations.}
        \label{FIG:Typical_evolution_of_perturbation} 
    \end{center}
\end{figure}

\begin{figure}[!ht]
    \begin{center}
        \includegraphics[scale=0.275]{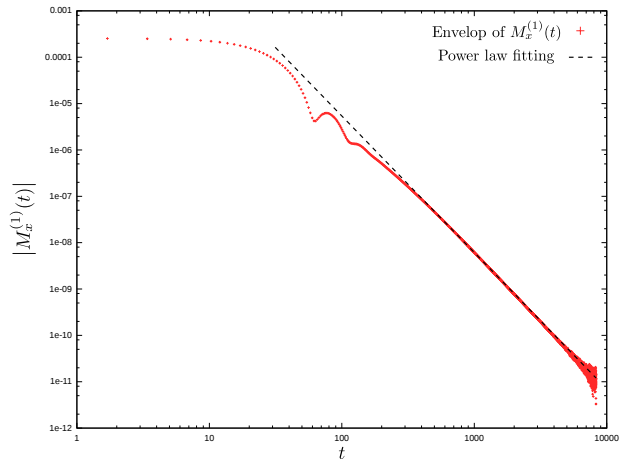}
        \caption{(color online) Temporal evolution of the
          $M_x^{(1)}(t)$ envelop for an initial condition perturbed by
          a cosine, as
          in~(\ref{Eq:Thermal_equilibrium_plus_Cosine_perturbation}). Red
          points: numerical simulation using the weighted particles code with
          $T=0.1$, $a=0.1$ and $N=10^{9}$. The initial weighted points
          are equally distributed in $]-\pi,\pi]\times[-3,3]$. 
          Black dashed line: power law fitting from $t=600$ to $6000$ ($\propto t^{-2.95}$).}
        \label{FIG:envelop_decay_of_Mx_cosine}
    \end{center}
\end{figure}

To estimate the perturbed magnetization $M_x^{(1)}(t)$, we subtract
from $M_x(t)$ its long time average.  A typical temporal evolution of
$M_x^{(1)}(t)$ is shown in
Fig.\ref{FIG:Typical_evolution_of_perturbation}. Finite size effects,
which are visible on the curve for $N=10^7$ points, may prevent the
study of the asymptotic behavior, so that it is usually necessary to use a very 
large number of points (see the curve for $N=10^8$ points).
We fit the envelop of the decaying $M_x^{(1)}(t)$ curve by a
power-law, using the least square method (see figure
(\ref{FIG:envelop_decay_of_Mx_cosine})). We find an exponent $-2.95$,
which is in very good agreement with the
prediction~(\ref{Eq:HMF_Mx_decay_prediction}) .

\subsection{Sine perturbation}
\label{sec:sine}

We consider now a sine perturbation of the thermal equilibrium: 
\begin{equation}
    \label{Eq:Thermal_equilibrium_plus_Sine_perturbation}
    f_0(\hmfx,p)+f_1(\hmfx,p) 
    =\mathcal{N}_ae^{-\beta\left(p^2/2-M_0\cos\hmfx\right)}\left(1+a\sin\hmfx\right).
\end{equation}
In this case, the $(\hmfx,p) \to (-\hmfx,-p)$ symmetry is broken, and we have
to compute the evolution of the whole $\mu$ space. A slow rotating
motion of the magnetization appears, which makes it difficult to
define an asymptotic average value of $M_x^{(1)}(t)$ and $M_y^{(1)}(t)$. Rather,
we used a running average to eliminate the rotation effect
(see \ref{sec:extraction}).
Finally, we obtain curves similar to the case of a cosine
perturbation to confirm the prediction
(\ref{Eq:HMF_Mx_decay_prediction}) and
(\ref{Eq:HMF_My_decay_prediction}).

\begin{figure}[!ht]
\begin{center}
    \begin{tabular}{cc}
        \includegraphics[scale=0.275]{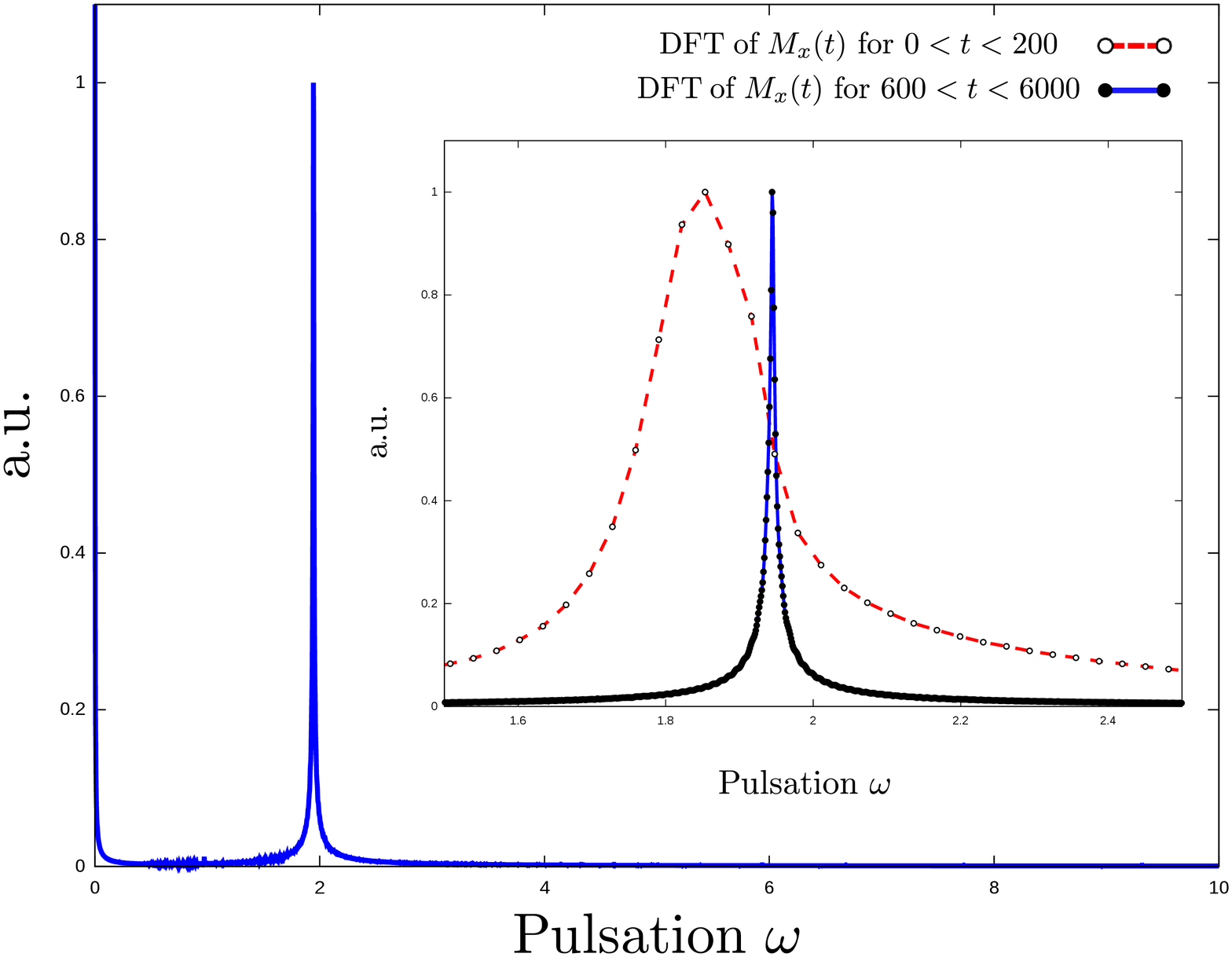}\\
        \includegraphics[scale=0.275]{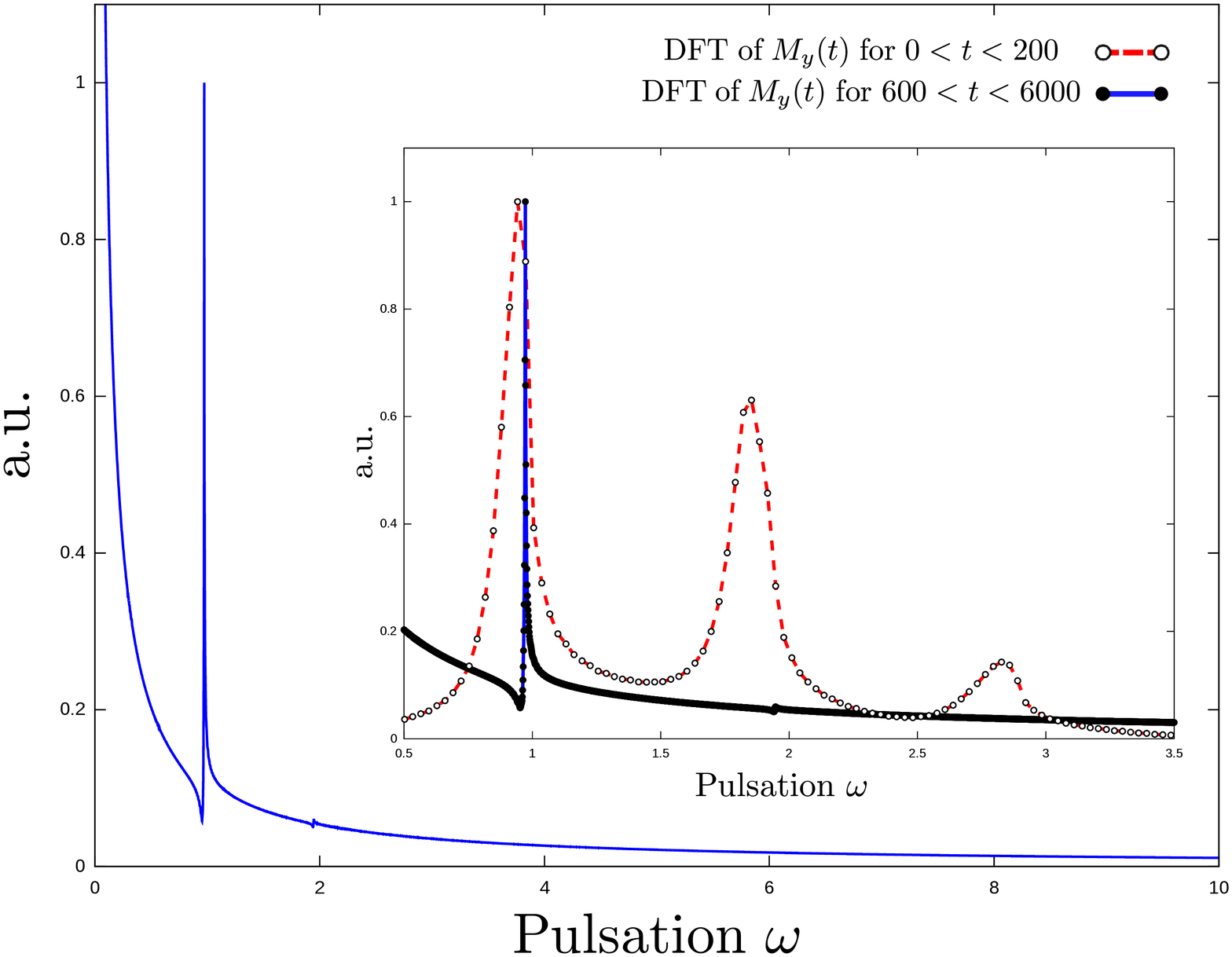}
    \end{tabular}
    \caption{(color online) Power spectra of the $M_x^{(1)}(t)$ (upper panel) and
      $M_y^{(1)}(t)$ (lower panel) for an initial condition perturbed by a sine, as in
      (\ref{Eq:Thermal_equilibrium_plus_Sine_perturbation}). The numerical
      simulation was done using the weighted particles
      code with $T=0.1$, $a=0.1$ and $N=10^{9}$.
      The initial weighted points are equally
      distributed in $]-\pi,\pi]\times[-3,3]$.  Each inset includes a magnification
      around the maxima and the short time power spectrum. The maxima of
      the long time power spectra of $M_x^{(1)}(t)$ and $M_y^{(1)}(t)$ are
      respectively $1.944$ and $0.973$.}
      \label{FIG:power_spectrum_of_Mx_My_Sine}
\end{center}
\end{figure}

We then fit the envelops of $M_x^{(1)}(t)$ and $M_y^{(1)}(t)$ with 
power laws. The exponents are $-2.94$ and $-1.77$, to be compared with
the predicted $-3$ and $-2$, see~(\ref{Eq:HMF_Mx_decay_prediction})
and~(\ref{Eq:HMF_My_decay_prediction}).  The relative error for
$M_y^{(1)}(t)$ exponent is close to $10\%$.  Different explanations
are possible: the fit range does not completely lie in the asymptotic
regime and/or there are numerical errors.
\begin{figure}[!ht]
    \begin{center}
        \begin{tabular}{c}
            \includegraphics[scale=0.275]{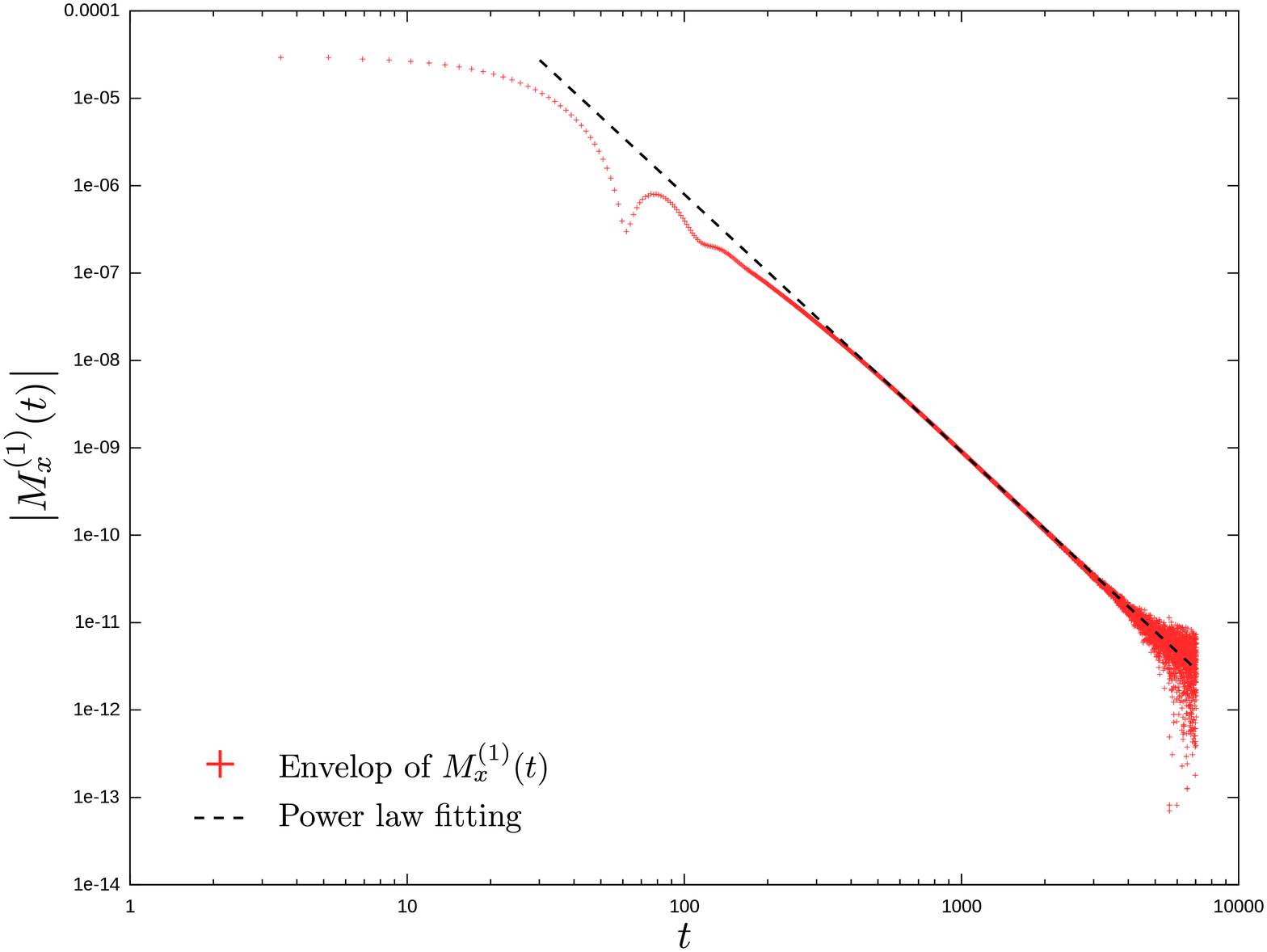} \\
            \includegraphics[scale=0.275]{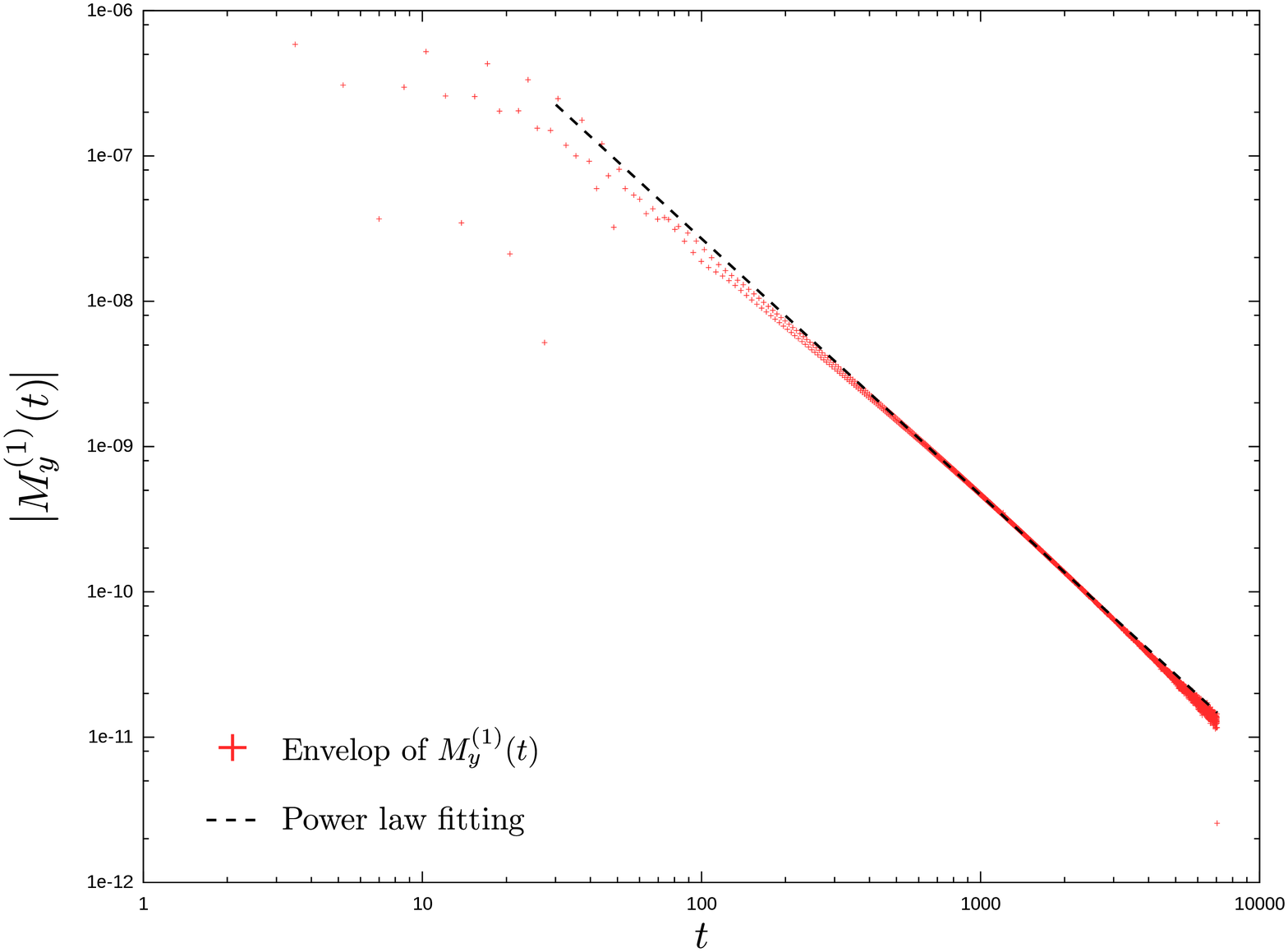}
        \end{tabular}
        \caption{(color online) Temporal evolutions of the
          $M_x^{(1)}(t)$ (upper panel) and $M_y^{(1)}(t)$ (lower panel) envelops
          for an initial condition perturbed by a sine, as in
          (\ref{Eq:Thermal_equilibrium_plus_Sine_perturbation}). Red
          points: numerical simulation using the weighted particles code with
          $T=0.1$, $a=0.1$ and $N=10^{9}$. The initial weighted points
          are equally distributed in $]-\pi,\pi]\times[-3,3]$.
          Black dashed line: power law fit from $t=600$ to $6000$.  We obtain
          an exponent equal to $-2.94$ for $M_x^{(1)}(t)$ and $-1.77$
          for $M_y^{(1)}(t)$.}
        \label{FIG:envelop_decay_of_Mx_My_Sine}
    \end{center}
\end{figure}
(\ref{Eq:HMF_Mx_decay_prediction}) and
(\ref{Eq:HMF_My_decay_prediction}) predict an oscillating decay of the
perturbations $M_x^{(1)}(t)$ and $M_y^{(1)}(t)$, with frequency
respectively $\omega_0$ and $2\omega_0$.
On Fig.~\ref{FIG:power_spectrum_of_Mx_My_Sine}, we plot the power spectra
of $M_x^{(1)}(t)$ and $M_y^{(1)}(t)$. We observe that indeed in both
cases a single frequency is selected in the long time regime, with
numerical values respectively $1.944$ and $0.973$. This is in almost
perfect agreement with the theoretical prediction $2\omega_0=1.945$
and $\omega_0=0.972$.

Figure \ref{FIG:power_spectrum_of_Mx_My_Sine} also allows to observe
the cross-over between two different dynamics explored by the system:
the short time evolution is driven by the Landau pole contribution,
and the asymptotic behavior by (\ref{Eq:HMF_Mx_decay_prediction})
and~(\ref{Eq:HMF_My_decay_prediction}).  Indeed, as shown
in~\cite{BOY10}, it is possible to compute the dominant Landau pole
for the parameters of Fig.~\ref{FIG:power_spectrum_of_Mx_My_Sine};
one finds a frequency $\rm Re(\omega_L)\simeq 1.8$.
This is in good agreement with
the short time power spectrum of $M_x(t)$, which is maximal around
$\omega=1.85$.
We therefore conclude that Landau damping occurs
in the short time regime, before the algebraic decay dominates.
The peak position at short time in the $M_{y}$'s spectrum
may be a signature of the growing $\omega_{0}$ peak
in a long time regime, since it is close to $\omega_{0}$.
It might also be related to a Landau pole associated to $M_y$:
since the matrix $\Lambda-F(\omega)$ is diagonal, Landau poles 
for $M_{x}$ and $M_{y}$ may be different.
Finally notice that the power spectrum divergence
close to $\omega=0$ is only due to the rotation of the
magnetization.

\section{Conclusion}
\label{sec:conclusion}

We have investigated the asymptotic dynamics of perturbations around
stable non homogeneous backgrounds in spatially one-dimensional Vlasov
equations.  The dispersion relation of the linearized Vlasov equation
has poles in the lower half of the complex plane and logarithmic
branch points on the real axis.  The poles yield exponentially
decaying contributions: this is a form of Landau damping.  The branch
points yield algebraically decaying contributions. If the dominant
branch point is related to a minimum of the potential created by the
stationary state, 
the leading order of the potential is quadratic with respect to position
around the minimum,
and unless some special cancellation occurs, the
perturbation potential behaves asymptotically as $e^{i\omega_0t}/t^2$,
where $\omega_{0}$ is the harmonic frequency of the potential well. We
expect this situation to happen in many cases of interest.

We have tested the theory on the HMF model by performing
$N$-body simulations which correspond to the full Vlasov equation
in the large $N$ limit; these simulations used the weighted particles code.
The exponent and frequency of the decay have been confirmed by these simulations, including in one case where a special symmetry imposes a decay as $e^{i2\omega_{0}t}/t^3$.

This summarizes in the following scenario for decaying perturbations
around stable non homogeneous background in one-dimensional systems:\\
i) The perturbation potential first roughly behaves as an oscillating decaying exponential, with the frequency and decay rate related to the poles of the dispersion relation, as usual Landau damping on homogeneous backgrounds.\\
ii) After this transient, the algebraic decay sets in, and the frequency changes; the decay exponent and the new frequency are now governed by the dominant branch point singularity on the real axis of the dispersion relation. 

These results prompt several questions, which this work does not answer. 
First, when the stationary distribution has a compact support in action variable, 
the edge of the support may create a singularity stronger than the bottom of the potential well. In some cases, one would then expect a $1/t$ asymptotic decay, with a frequency
corresponding to the action at the edge of the support. However, a stationary distribution with compact support may also sustain purely oscillatory modes~\cite{Mathur90}, which do not decay at all. We have not been able to find a stationary state with compact support and no oscillatory mode to test the possibility of a $1/t$ decay.
Second, the local extrema of the function $\Omega(J)$ may also create a different type of singularity, and thus modify the asymptotic behavior of a perturbation.
Recently, Bouchet and Morita have studied a similar situation in the 
context of the 2D Euler equation, unveiling the phenomenon of ``vorticity depletion'' close to these local extrema~\cite{BouchetMorita09}. The possibility of a similar behavior for the Vlasov equation is an open question.
Third, we picked up the slowest decaying contribution
among Fourier modes with respect to the angle variable,
and we have not investigated if the sum over the infinite number of Fourier modes
affects the asymptotic decay.

It is also of primary interest to understand the asymptotic behavior of a perturbation in a three dimensional setting. When the potential created by the stationary state is integrable, angle-action variables can be defined, and the method 
used in this article is viable: one would have to study the singularities of the dispersion relation in this case. In a situation where the potential created by the stationary state is not integrable, the strategy would fail. 

Finally, our numerical study required to compute a Vlasov evolution with good precision for a long time, which raises difficulties.
To overcome them, we have introduced a particle method, the weighted particles approach. It proved particularly well-suited for our purpose, and it would be interesting to investigate the reasons for this.

\ack
The authors thank Freddy Bouchet and Bruno Marcos for helpful discussions.
This work is supported by the Ministry of Educations, Science, Sports and Culture,
Grant-in-Aid for Young Scientists (B), 19760052 and by the ANR-09-JCJC-009401
 INTERLOP project. Computations have been partially performed on the
``M\'esocentre SIGAMM'' machine, hosted by the Observatoire de la
C\^ote d'Azur.

\appendix
\section{Estimate of $a_{k}(t)$}
\label{sec:estimate}

\begin{figure}[h]
    \centering
    \includegraphics[width=7cm]{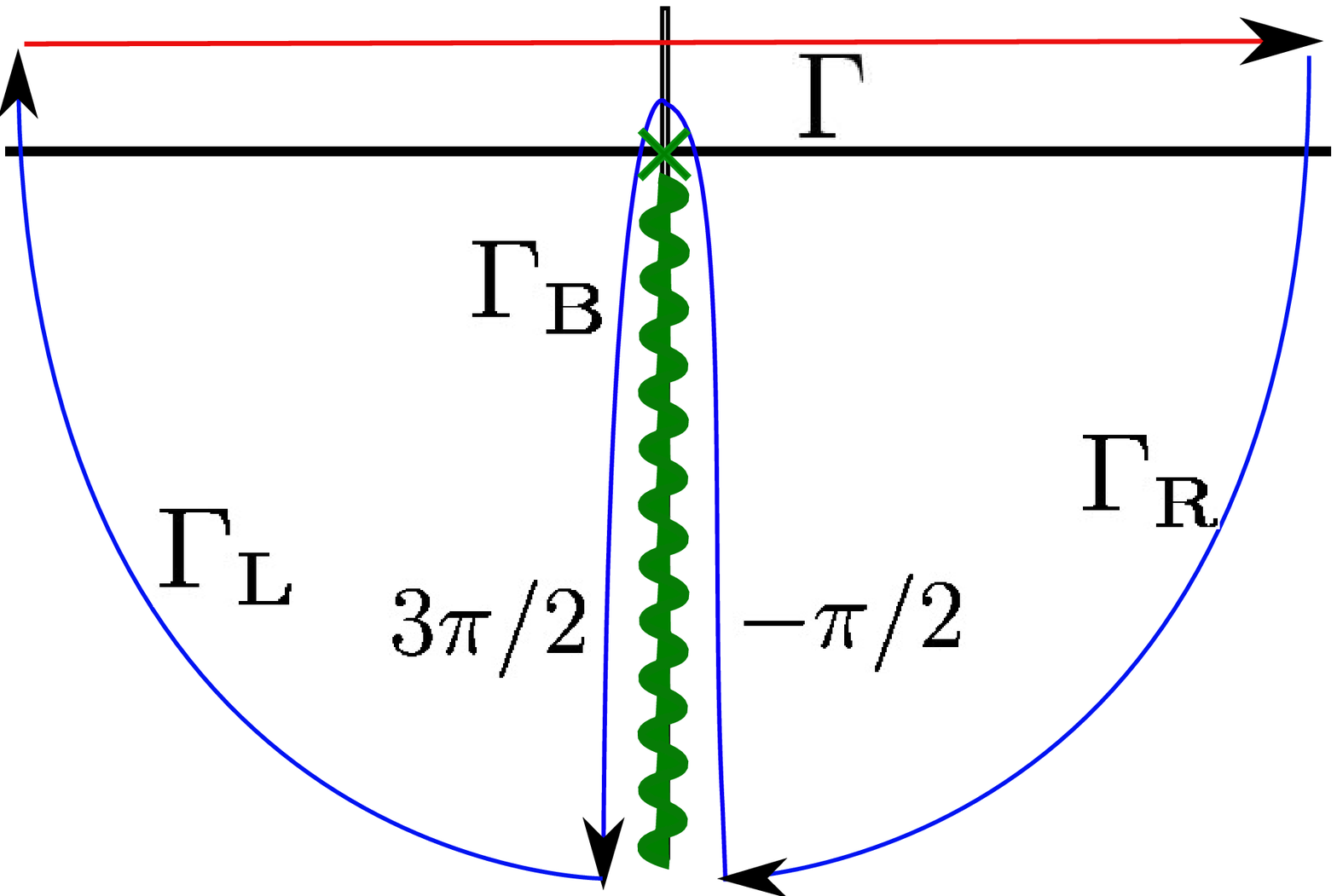}
    \caption{(color online) Modification of the Bromwich contour $\Gamma$
      to the closed path $\Gamma+\Gamma_{\mbox{R}}+\Gamma_{\mbox{B}}+\Gamma_{\mbox{L}}$
      in the complex $z$ plane.
      The wavy line represents a branch cut associated
      to the logarithmic singularity at $z=0$,
      which is assumed as the unique singularity.
      The numbers $3\pi/2$ and $-\pi/2$ are $\arg(z)$
      on the present Riemann sheet.}
    \label{fig:mod-bromwich}
\end{figure}

Each function $\tilde{a}_{k}(\omega)$ has singularities of
$z^\nu \ln z$ type at $\omega=m\omega_{a}$ for all $m$. It may be written as
\begin{equation}
\tilde{a}_{k}(\omega)=\sum_{m} A_{m}(\omega-m\omega_{a})
(\omega-m\omega_{a})^{\nu_{m}}\ln(\omega-m\omega_{a}) +B(\omega)
\label{eq:a-expansion}
\end{equation}
where the $A_{m}(z)$ are supposed to be analytic, and the possible singularities of $B$
at $\omega=m\omega_{a}$ are weaker than $z^{\nu_m}\ln z$. Isolating one
term in the sum over $m$, we have to compute the following inverse Laplace 
transform, dropping the index $m$ for simplicity:
\begin{eqnarray}
    \label{eq:a-estimate}
         \int_{\Gamma} A(z-m\omega_{a}) (z-m\omega_{a})^{\nu} \ln(z-m\omega_{a}) e^{-izt} dz
\nonumber \\
         = e^{-im\omega_{a}t} \int_{\Gamma} A(z) z^{\nu}\ln(z) e^{-izt} dz. 
\end{eqnarray}
We assume that $A(z)$ rapidly decreases for $|z|\to\infty$.
We take a branch cut in the lower half of the imaginary axis
as shown in Fig.~\ref{fig:mod-bromwich}.

To compute the integral over the Bromwich contour $\Gamma$,
we add the paths $\Gamma_{\mbox{R}},\Gamma_{\mbox{B}}$ and $\Gamma_{\mbox{L}}$
to the Bromwich contour $\Gamma$, and make a closed path.
From the assumption no singularity is enclosed by the closed path, and hence
\begin{equation}
    \label{eq:closed-path}
    \int_{\Gamma+\Gamma_{\mbox{R}}+\Gamma_{\mbox{B}} +\Gamma_{\mbox{L}}}
    A(z) z^{\nu} \ln(z) e^{-izt} dz = 0.
\end{equation}
Thanks to the factor $e^{-izt}$ of the integrand,
contributions from $\Gamma_{\mbox{R}}$ and $\Gamma_{\mbox{R}}$ vanish, and hence
\begin{eqnarray}
    \label{eq:Gamma-GammaB}
         \int_{\Gamma} A(z) z^{\nu} \ln z e^{-izt} dz
        = \int_{-\Gamma_{\mbox{B}}} A(z) z^{\nu} \ln(z) e^{-izt} dz \\
         = \int_{\infty}^{0} A(re^{i 3\pi/2}) (re^{i 3\pi/2})^{\nu} \ln(re^{i 3\pi/2})        e^{-ire^{i 3\pi/2}t} e^{i 3\pi/2}dr \nonumber \\
         + \int_{0}^{\infty} A(re^{-i\pi/2}) (re^{-i\pi/2})^{\nu} \ln(re^{-i\pi/2}) e^{-ire^{-i\pi/2}t} e^{-i\pi/2}dr
\end{eqnarray}
Due to the branch cut, we have to distinguish $A(re^{i 3\pi/2})$ from $A(re^{-i \pi/2})$,
and we denote them by $A_{\mbox{L}}(r)$ and $A_{\mbox{R}}(r)$ respectively.
The integral is hence written as
\begin{eqnarray}
    \label{eq:Gamma-branch}
         \int_{\Gamma} A(z) z \ln z e^{-izt} dz  
        & = & i(-i)^{\nu} \int_{0}^{\infty} [ A_{\mbox{L}}(r) - A_{\mbox{R}}(r) ] r^{\nu} \ln(r) e^{-rt} dr \nonumber \\
        && - (-i)^{\nu} \frac{\pi}{2} \int_{0}^{\infty} [ 3A_{\mbox{L}}(r) + A_{\mbox{R}}(r)] r^{\nu} e^{-rt} dr.
\end{eqnarray}
The two functions $A_{\mbox{L}}(r)$ and $A_{\mbox{R}}(r)$ coincide
in the limit $r\to 0$. Around the singularity $r=0$, we can therefore estimate
\begin{eqnarray}
    \label{eq:A-estimate}
        & A_{\mbox{L}}(r) - A_{\mbox{R}}(r) \propto r, \\
        & 3A_{\mbox{L}}(r) + A_{\mbox{R}}(r) \propto r^{0}.
\end{eqnarray}
Using a scaling of the variable as $y=rt$,  the integral is expressed by
\begin{eqnarray}
    \label{eq:integral-final}
        \int_{\Gamma} A(z) z \ln z e^{-izt} dz 
        & = \frac{C_{0}}{t^{\nu+2}} \int_{0}^{\infty} y^{\nu+1} (\ln y-\ln t) e^{-y} dy \nonumber \\
        &+ \frac{C_{1}}{t^{\nu+1}} \int_{0}^{\infty} y^{\nu} e^{-y} dy.
\end{eqnarray}
The first term of the right-hand-side yields $t^{-\nu-2}$ decay and $t^{-\nu-2}\ln t$ decay,
and the second term $t^{-\nu-1}$ decay.
Returning to (\ref{eq:a-estimate}), we obtain 
$e^{-im\omega_{a}t}/t^{\nu+1}$ as the slowest decay.
From (\ref{eq:a-expansion}), we now see that the asymptotic 
decay of $a_k(t)$ is governed by the smallest $\nu_{m}$.

\section{Weighted particles code}
\label{sec:semi-vlasov-code}
The $N$-body simulations are performed by the weighted particles code.
In a standard $N$-body simulations, we would prepare $N$ initial positions and momenta
by drawing random numbers according to a given initial distribution.
An initial condition prepared in this way has fluctuations of order $O(1/\sqrt{N})$.
In the weighted particles code, we prepare $N$ initial positions and momenta
as lattice points of a square lattice, and give to each lattice point a weight proportional to 
the initial distribution we want to sample. The concrete algorithm is as follows.

We span the $\mu$ space with a regular lattice, having $N$ points.
The upper and lower boundaries $\pm p_{\mbox{max}}$ for the lattice in
the $p$-direction must be set such that the initial distribution
$f(\hmfx,p)$ for $|p|>p_{\mbox{max}}$ is negligible.  Let
$i=1,\cdots,N$ denote the lattice points, and $(\hmfx_{i},p_{i})$
their coordinates.  We assign the weight $w_{i}=C f(\hmfx_{i},p_{i})$
to each lattice point, where the constant $C$ is defined by the
normalization
\begin{displaymath}
    \sum_{i=1}^{N} w_{i} = 1.
\end{displaymath}
We put one particle on each lattice point, 
and the particles move on $\mu$ space following by the canonical equations of motion
for the HMF model
\begin{displaymath}
    \dot{\hmfx}_{i}(t) = p_{i}(t), \quad
    \dot{p}_{i}(t) = - \bar{M}_{x}(t)\sin\hmfx_{i}(t) + \bar{M}_{y}(t)\cos\hmfx_{i}(t),
\end{displaymath}
where the suffix $i$ runs from $1$ to $N$, and $\bar{M}_{x}(t)$ and $\bar{M}_{y}(t)$
are defined by
\begin{equation}
    \label{eq:MxMy-weight}
    \bar{M}_{x}(t) = \sum_{i=1}^{N} w_{i} \cos\hmfx_{i}(t), \quad
    \bar{M}_{y}(t) = \sum_{i=1}^{N} w_{i} \sin\hmfx_{i}(t).
\end{equation}
We note that the lattice is used only to define the weight $w_{i}$,
and to set the initial condition for the $i$-th particle as 
$(\hmfx_{i}(0),p_{i}(0))=(\hmfx_{i},p_{i})$.
It is worth stressing that particles are not fixed at lattice points,
but move in the whole $\mu$ space, keeping their initially assigned weights $w_{i}$.
The magnetization $(M_{x},M_{y})$ is computed using the evolving positions
and the initially fixed weight as (\ref{eq:MxMy-weight}).

In a semi-Lagrangian code \cite{buyl10}, the distribution is defined
at fixed lattice points.  To evolve the distribution over a time step
$\Delta t$, one computes the inverse temporal evolution of a particle
during $\Delta t$ with initial condition given by the lattice
$(\hmfx_{i}(0),p_{i}(0))=(\hmfx_{i},p_{i})$.  Integrating the Vlasov
equation, one gets $f(\hmfx_{i},p_{i},\Delta t)=f(\hmfx_{i}(-\Delta
t),p_{i}(-\Delta t),0)$.  The point $(\hmfx_{i}(-\Delta t),p_{i}(-\Delta t))$
does not coincide with a lattice point generally, so that
the value of the distribution at this point is obtained by
interpolation.  The semi-Lagrangian code thus requires three
discretization parameters: the time step $\Delta t$, and two spatial
and velocity discretization parameters $\Delta\hmfx,\Delta p$. In
order to insure numerical stability, the time step must become small
as spatial and velocity discretizations become small.  As a result, we
need a small time step for computations with high spatial resolution,
which results in heavy computations to reach the long time regime.

In the weighted particles code,
we may set the $\Delta\hmfx$ and $\Delta p$ discretizations
independently of the time step $\Delta t$,
and hence the computational burden may be reduced by taking a rather large $\Delta t$,
still giving a good enough accuracy.

The weighted particles code has further advantages
against the semi-Lagrangian code:\\
(1) The Vlasov equation has an infinite number of conserved quantities
which are $\int (f(\hmfx,p))^{l} d\hmfx dp$ for $l\in\mathbb{N}$,
but it is known that the semi-Lagrangian code cannot preserve them for $l\geq 2$.
In the weighted particles code, the weight $w_{i}$ is a fixed value
and hence all such quantities, approximated by $\sum_{i} (w_{i})^{l}$,
are preserved exactly.
(2) The semi-Lagrangian code uses interpolation.
It is a delicate step to obtain the temporal evolution of the distribution function,
and the accuracy of the code depends on the algorithm of interpolation.
The weighted particles code does not require any interpolation.

Let us remark that the weighted particles algorithm is highly parallelizable and
its convenient structure makes it possible to use a lot of parallelization
methods. In particular, it allows to take advantage of the available computer architecture, be it a cluster with distributed memory or shared memory. In the case of a distributed memory,
the mean field property allows to restrict the communication between node to the magnetization, which can be computed piece by piece on each node.
It is thus possible to compute the long time evolution of the
system for a very large number of weighted particles.

One of the disadvantages of the weighted particles codes is that this code
cannot compute the temporal evolution of the distribution directly. We
can obtain a coarse-grained distribution, but its resolution is lower
than the initially given lattice.  Another disadvantage may be the
limitation of objects for which the weighted particles code works well.  The
weight $w_{i}$ on a lattice point corresponds to set several particles
with the same initial condition. If these particles were given
slightly different initial conditions, they would eventually separate
as time goes by.  In the weighted particles code, they remain
together. Consequently, weighted particles code might have to be improved
if it is to be used in order to observe more drastic changes of the
distribution function, such as violent relaxation from a waterbag
initial state to a Lynden-Bell quasi-stationary state. 
Further investigations are needed to understand why the weighted particles code seems to work so well in our case. 
The numerical tests and theoretical arguments given in \cite{WollmanOzizmir96} may be a first step in this direction.

\section{Extraction of a rotating part}
\label{sec:extraction}

\begin{figure}[!ht]
     \begin{center}
         \begin{tabular}{cc}
             \includegraphics[scale=0.275]{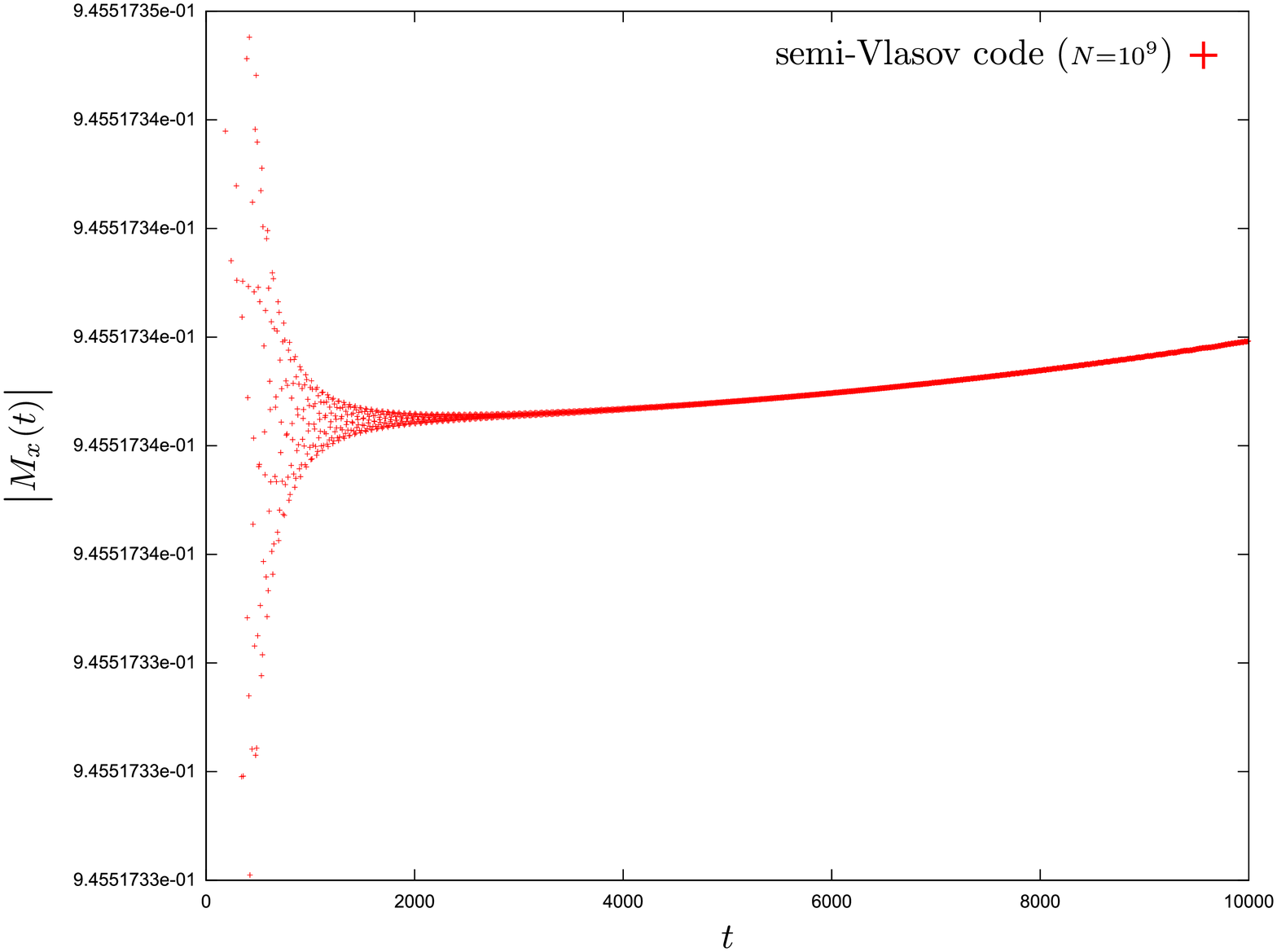}\\
             \includegraphics[scale=0.275]{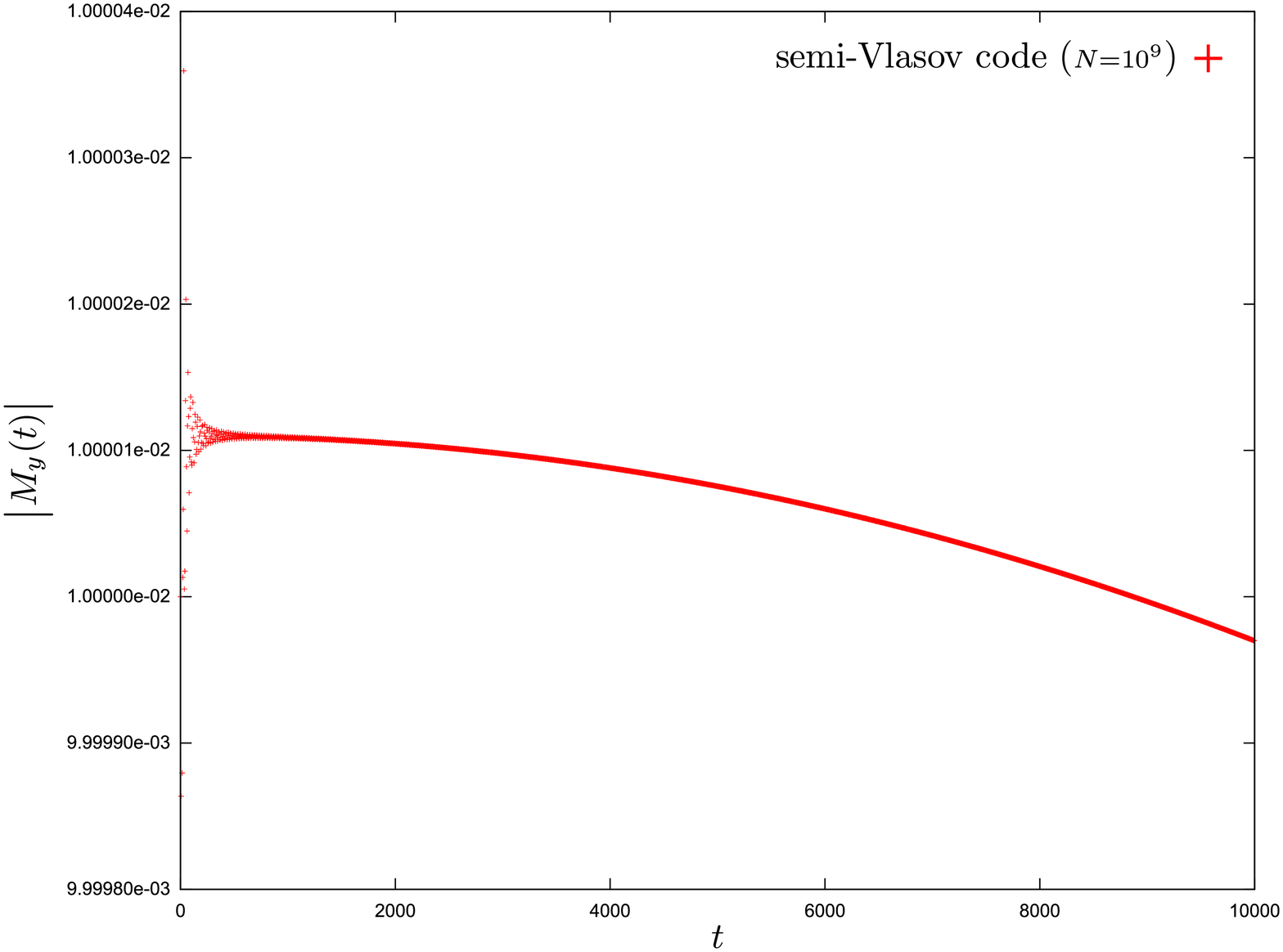}
         \end{tabular}
         \caption{Slow rotation of $M_x^{(1)}(t)$ (upper panel) and $M_y^{(1)}(t)$ (lower panel)
           for an initial condition perturbed by a sine, as in
           (\ref{Eq:Thermal_equilibrium_plus_Sine_perturbation}). The
           numerical simulation was done using the weighted particles code
           with $T=0.1$, $a=0.1$ and $N=10^{9}$. The initial weighted
           points $w_i$ are equally distributed in $]-\pi,\pi]\times[-3,3]$.}
          \label{FIG:Rotation_of_Mx_My_for_sine_perturbation}
      \end{center}
 \end{figure}

\begin{figure}
    \begin{center}
        \begin{tabular}{c}
            \includegraphics[scale=0.18]{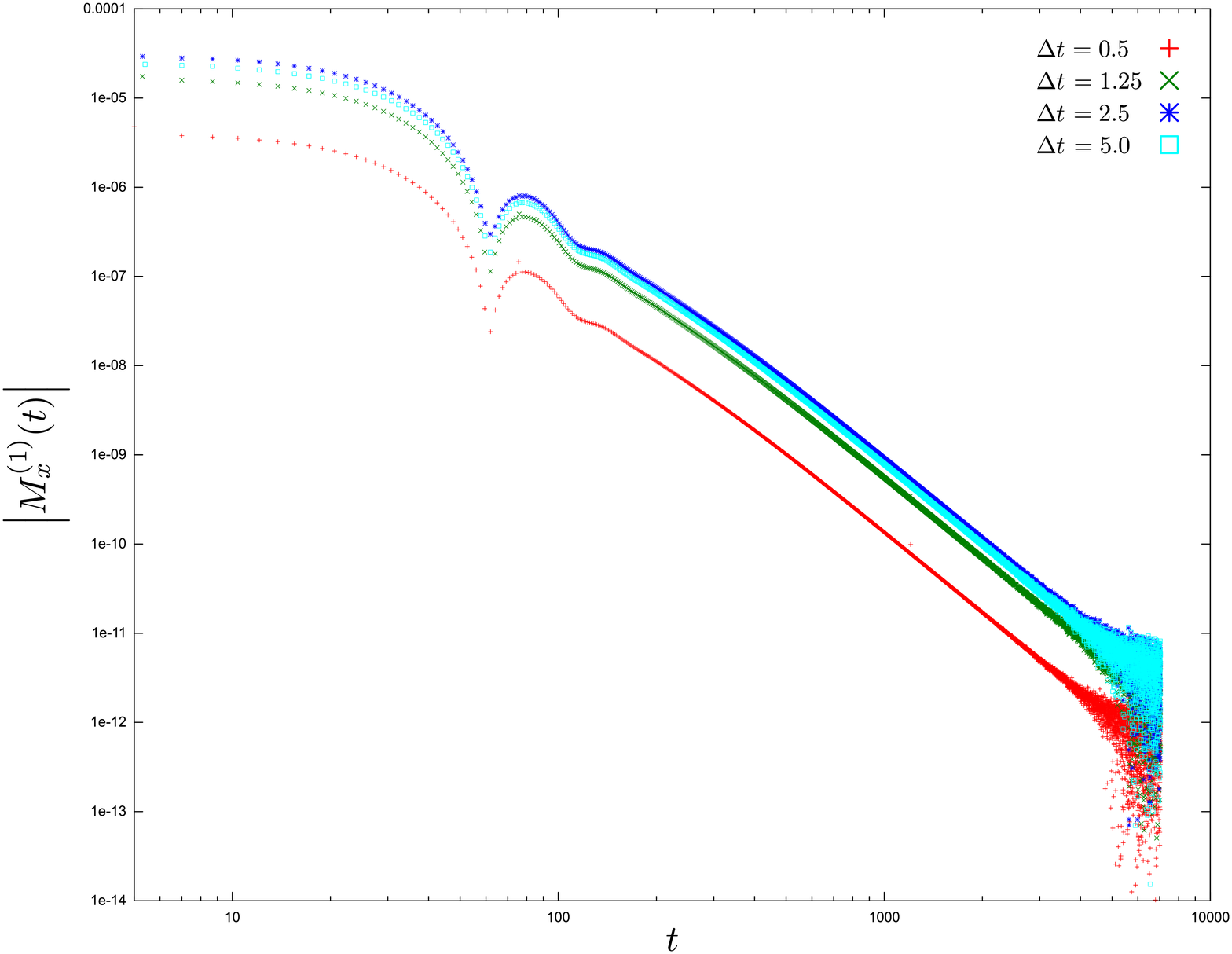}\\
            \includegraphics[scale=0.18]{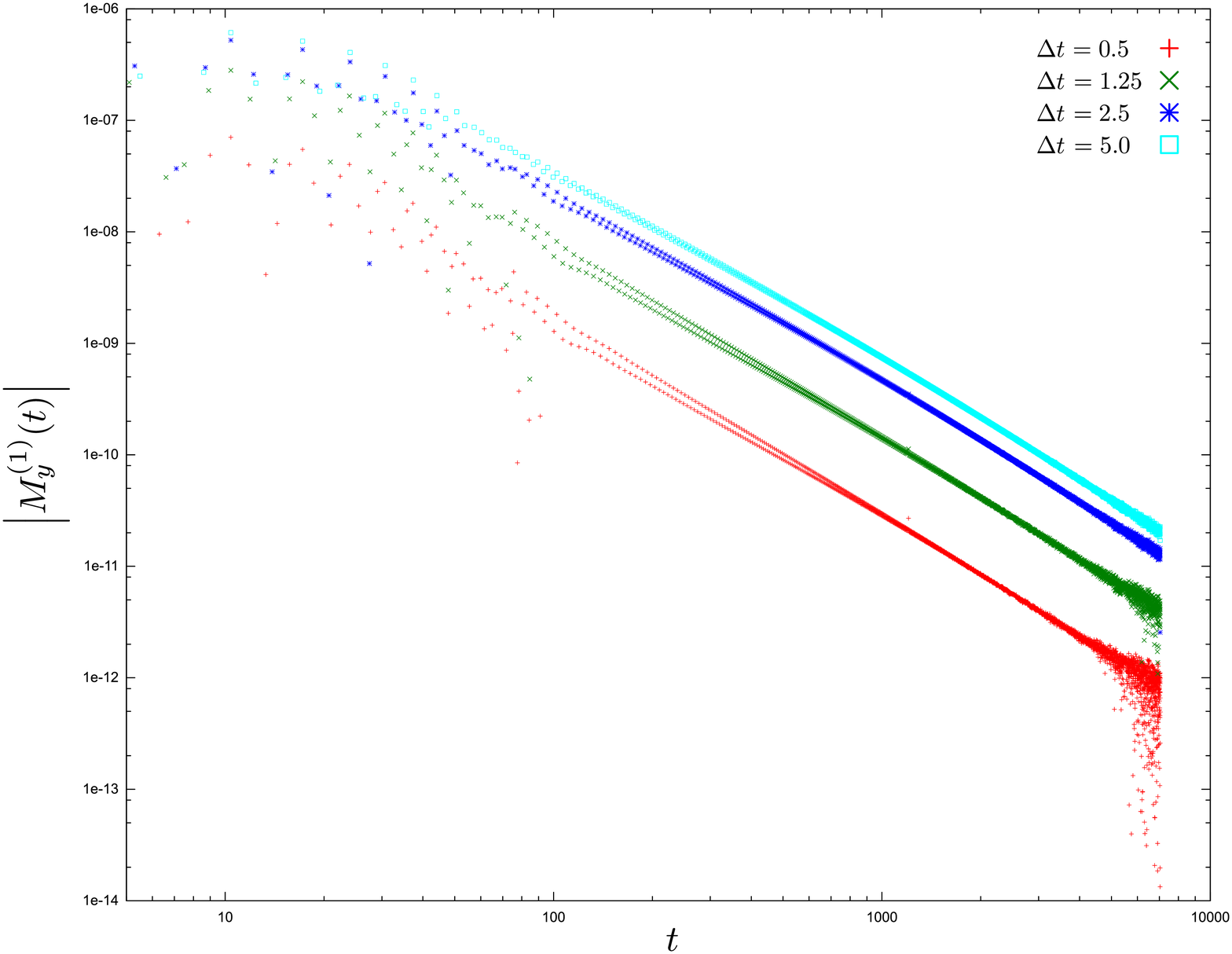}
        \end{tabular}
        \caption{(color online) Temporal evolution of the
          $M_x^{(1)}(t)$ and $M_y^{(1)}(t)$ envelop for an initial
          condition perturbed by a sine, as in
          (\ref{Eq:Thermal_equilibrium_plus_Sine_perturbation}),
          using different window sizes for the running average. The
          numerical simulation was done using the weighted particles code
          with $T=0.1$, $a=0.1$ and $N=10^{9}$. The initial weighted
          points are equally distributed in $]-\pi,\pi]\times[-3,3]$.}
        \label{FIG:envelop_decay_test_of_Mx_My_Sine}
    \end{center}
\end{figure}

When we use a asymmetric perturbation such as the one given in (\ref{Eq:Thermal_equilibrium_plus_Sine_perturbation}), we observe a small rotation of the $x$ and $y$ magnetizations $M_x(t)$ and $M_y(t)$ (see Fig.~\ref{FIG:Rotation_of_Mx_My_for_sine_perturbation}). 
 In this case, it is not possible to directly define $M_x^{(1)}(t)$
 and $M_y^{(1)}(t)$ by substracting the long time average.  Our method
 is then to use a running average. We define $M_x^{(1)}(t)$ and
 $M_y^{(1)}(t)$ such that
\begin{equation}
    M_x^{(1)}(t)=M_x(t)- \frac{1}{2\Delta t}\int_{t-\Delta t}^{t+\Delta t}M_x(t')dt',
\end{equation}
and
\begin{equation}
    M_y^{(1)}(t)=M_y(t)- \frac{1}{2\Delta t}\int_{t-\Delta t}^{t+\Delta t}M_y(t')dt'.
\end{equation}
In order to compute the right exponent of a power law fit, we have to
choose the parameter $\Delta t$.  However our different tests (see Fig.~
\ref{FIG:envelop_decay_test_of_Mx_My_Sine}) show that modifying $\Delta t$
does not change much the result.  We have taken $\Delta t=5$ for Fig.~
\ref{FIG:envelop_decay_of_Mx_My_Sine}. This is roughly the time needed
to observe one oscillation of $M_y(t)$, and two oscillations of $M_x(t)$.

\end{document}